\documentclass[12pt]{iopart}

\usepackage{iopams}

\expandafter\let\csname equation*\endcsname\relax

\expandafter\let\csname endequation*\endcsname\relax
\usepackage{comment}
\usepackage{amsmath}
\usepackage{cite} 
\usepackage{tikz}
\usepackage{graphicx}
\usepackage{soul}
\usetikzlibrary{matrix,shapes,arrows,positioning,chains} 

\usepackage{hyperref}
\usepackage{amsmath}
\usepackage{amsfonts}
\usepackage{amssymb,latexsym,mathrsfs}
\usepackage{graphicx} 
\usepackage{float}

\begin{document}

\title[First detection probability in quantum resetting via random projective measurements]{First detection probability in quantum resetting via random projective measurements}

\author{Manas Kulkarni$^1$ and Satya N. Majumdar$^2$}

\address{$^1$ International Centre for Theoretical Sciences, Tata Institute of Fundamental Research, Bengaluru -- 560089, India\\
$^2$ LPTMS, CNRS, Univ. Paris-Sud, Universit\'e Paris-Saclay, 91405 Orsay, France }
\ead{manas.kulkarni@icts.res.in, satya.majumdar@u-psud.fr}


\begin{abstract}

We provide a general framework to compute the probability distribution $F_r(t)$ of the first detection 
time of a `state of interest' in a generic quantum system subjected to random projective measurements. In 
our `quantum resetting' protocol, resetting of a state is not implemented by an additional classical 
stochastic move, but rather by the random projective measurement. We then apply this general framework to 
Poissonian measurement protocol with a constant rate $r$ and demonstrate that exact results for $F_r(t)$ 
can be obtained for a generic two level system. Interestingly, the result depends crucially on the 
detection schemes involved and we have studied two complementary schemes, where the state of interest 
either coincides or differs from the initial state. We show that $F_r(t)$ at short times vanishes 
universally as $F_r(t)\sim t^2$ as $t\to 0$ in the first scheme, while it approaches a constant as $t\to 
0$ in the second scheme. The mean first detection time, as a function of the measurement rate $r$, also 
shows rather different behaviors in the two schemes. In the former, the mean detection time is a 
non-monotonic function of $r$ with a single minimum at an optimal value $r^*$, while in the later, it is a 
monotonically decreasing function of $r$, signalling the absence of a finite optimal value. These general 
predictions for arbitrary two level systems are then verified via explicit computation in the 
Jaynes-Cummings model of light-matter interaction. We also generalise our results to non-Poissonian measurement protocols 
with a renewal structure where the intervals between successive independent measurements are distributed 
via a general distribution $p(\tau)$ and show that the short time behavior of 
$F_r(t)\sim p(0)\, t^2$ is universal as long as $p(0)\ne 0$. This universal $t^2$ law emerges from purely quantum dynamics that
dominates at early times.

\end{abstract}


\section{Introduction}

Stochastic resetting in classical systems has emerged as a major area of research, both theoretically and 
experimentally, with a large number of applications across disciplines (for recent reviews 
see~\cite{EMS20,GJ22,PKR22}). The basic idea behind stochastic resetting is as follows~\cite{EM11,EM12}. Consider 
any system evolving under its own natural dynamics (which can be stochastic or deterministic). This natural 
dynamics is interrupted at random times and then the system restarts afresh from a `reset' configuration. This 
reset configuration is typically the initial configuration, but it can be any other fixed 
configuration~\cite{EM12}. The time interval $\tau$ between two successive resets is chosen independently from a 
distribution $p(\tau)$. For example, if $p(\tau)= r\, e^{-r\,\tau}$ (where the resetting occurs at a constant rate 
$r$) it is called the Poissonian resetting~\cite{EMS20}, while if reset occurs periodically/stroboscopically with 
a period $T$, it is referred to as `sharp restart'~\cite{PKE_16,Reu_16}. There are two principal outcomes of this 
resetting dynamics~\cite{EM11}: (i) resetting breaks detailed balance manifestly and drives the system into a 
nonequilibrium stationary state (NESS) and (ii) resetting typically makes the search of a fixed target by a random 
search process more efficient. For instance, the mean first-pasage time to find a target in space by a diffusive 
searcher in any dimension gets minimized for a certain choice of the resetting rate $r^*$~\cite{EM11,EM_14}.  
These two paradigms have been tested in numerous theoretical 
models~\cite{EM_14,KMSS_14,GMS_14,MSS_15,CS_15,CM_15,PKE_16,Reu_16,MV_16,PR_17,CS_18,MMS_20,Bres_20,Pinsky_20,BMS_22,BLMS_23,BMS_23} 
and also in optical tweezer experiments in the recent past~\cite{TPSRR_20,BBPMC_20,FBPCM_21}.
  
The same question can be asked also for a quantum system where the natural dynamics is deterministic, provided by 
the unitary evolution governed by the Schr\"odinger equation. The state of the system at time $t$ is denoted by 
$|\psi(t)\rangle$. When this unitary evolution is interrupted at Poisson distributed random times and the state of 
the quantum system `restarts' from its initial state, the density matrix of the system evolves into a NESS with 
nonzero off-diagonal elements~\cite{MSM18} and this NESS has been investigated in various quantum 
systems~\cite{MSM18,RTLG18,PCML21,PCL22}. However, the resetting of the quantum state to $|\Psi(0)\rangle$ is 
typically implemented `by hand', i.e., by a classical stochastic move in these models. How this `quantum 
resetting' can be actually implemented without using classical stochastic moves is an interesting question by 
itself. The other interesting issue is to define an analogue of the classical first passage probability in a 
quantum system. One way to implement this `quantum resetting' is via standard projective 
measurement~\cite{WM_Book_2009,NC_Book_2010,J_Book_2014}. Suppose, we have a detector which tries to detect at 
`random times' if the quantum system is in a particular `state of interest' which we will denote by $|\psi_{\rm 
int}\rangle$. If the outcome is `yes', the state of the system immediately after the measurement attempt becomes 
$\hat{P}\, |\psi(t)\rangle$, where $\hat{P}= |\psi_{\rm int}\rangle \langle \psi_{\rm int}|$ denotes the 
projection operator into the state of interest $|\psi_{\rm int}\rangle$. In contrast, if the outcome is `no' the 
state of the system immediately after the measurement gets projected into $\hat{Q}|\psi(t)\rangle$, where 
$\hat{Q}=I-\hat{P}$ denotes the projection operator into the subspace of the Hilbert space that is complementary 
to the state of 
interest~\cite{KB06.1,KB06.2,VKB08,GVWW13,DDD15,DDDS15,FKB17,TBK18,YZTB19,LD19,SKR20,LYZB20,DBD21,DDG21,
DDG22,DG22,MCPL22,DCD23,SV23,YB23.1,YB23.2,LSC23}. 
Then the system evolves unitarily till the next measurement attempt and so on. These measurements can occur 
stroboscopically with period $T$ -- this is the analogue of `sharp restart' in classical systems. Similarly when 
the detection attempts take place at random epochs separated by an exponential distribution $p(\tau)= r\, e^{-r\, 
\tau}$, it is the analogue of Poissonian resetting in classical systems. Note that here the `quantum restart' is 
not implemented by hand using classical moves, but rather the projective measurement itself does the job of 
resetting to a new state after every measurement attempt. 

This projective measurement protocols also provide a 
natural analogue of the classical first passage probability, namely the first detection probability. For example, 
if the outcomes of successive measurements are $\{ {\rm no}, {\rm no},\ldots, {\rm no}, {\rm yes}\}$, then when 
`yes' appears for the first time, we say that the state is detected at that time. Clearly this first detection 
time is a random variable if the measurement times are also random and its probability distribution function (PDF) 
is the first detection probability~\cite{KB06.1, KB06.2,VKB08,GVWW13,DDD15,DDDS15,FKB17,TBK18,YZTB19,LD19,TMMBK20,KEZ21,ZBK21, 
DG22,TZ22}. Its first moment, namely the mean first detection time is thus a natural analogue of the mean first passage 
time in classical systems. The resetting rate in these quantum systems is just the rate of detection attempts, 
i.e., the rate of measurement. Note that this `quantum resetting' via projective measurements is also different 
from the effective stochastic resetting dynamics exhibited by non-Hermitian quasiparticles describing the growth 
of entanglement in a class of quantum systems coupled to a measurement device~\cite{TDFS22}.

Thus, in this paper, by `quantum resetting' we simply mean projective measurements only and do not invoke any 
additional `classical 
restart' moves by hand. We ask a simple and natural question: as in classical resetting systems, is there an optimal 
resetting rate, i.e., an optimal rate of measurement, that minimizes the mean first detection time in quantum resettings 
via projective measurements only? The existence of this optimal rate has been investigated recently in various quantum 
models, but usually in the presence of an additional `classical restart' move~\cite{YB23.1,YB23.2}. Here our goal is to 
investigate this optimal resetting rate in quantum systems subject only to random projective measurements 
via the Poissonian measurement protocol with rate $r$. We first provide a general 
framework to compute the first detection probability $F_r(t)$ in a quantum system with Poissonian protocol, and in particular, 
focus on generic two-level systems where explicit results can be obtained. 
We study two different detection schemes depending on whether the state of interest is different or identical to the initial 
state. We show the statistics of the first detection time is rather different in the two schemes.
For instance, $F_r(t)$ at short times vanishes 
universally as $F_r(t)\sim t^2$ as $t\to 0$ in the first scheme, while it approaches a constant as $t\to 
0$ in the second scheme. The mean first detection time, as a function of the measurement rate $r$, also 
shows rather different behaviors in the two schemes.
In the first scheme, the mean first detection time exhibits a unique minimum as a function of $r$, thus confirming 
the existence of an optimal measurement rate in this quantum setting. In contrast, for 
the second scheme, we show that the mean first detection time is a monotonically decreasing function of $r$, showing 
that there is no finite optimal measurement rate. This last fact in scheme 2 
was also noticed in Refs.~\cite{KEZ21, ZBK21}--where scheme 2 is usually referred to 
as the `return' problem of a quantum state to itself.

We then focus on a specific two level system,
the Jaynes-Cummings (JC) model which is a prototypical model mimicking light-matter
interaction~\cite{SZ97,C99,HR06,GSA13}. 
The JC model can be effectively mapped to a two-level system for a given fixed excitation
sector.  The JC model subjected to `classical stochastic resetting' (and not projective measurement) has been studied
very recently and certain observables such as the fidelity etc. have been computed~\cite{SV23}.
In this paper,  we show that the integrable structure of the JC model allows us
to compute explicitly the full PDF of the first detection time under projective
Poissonian measurement protocol. This exact solution is useful since we can then test and verify
all our general predictions valid for arbitrary two level systems.
Moreover, from the exact solution of the JC model we show that there exists another time scale
$t_m(r)$ characterzing the late time exponential decay of the PDF of the first detection time.
This time scale $t_m(r)$ is different from the standard mean first detection time which is
just the first moment of the PDF.
Physically, this time scale $t_m(r)$ represents the maximal 
waiting time to detect the state of interest beyond which there is no detection almost surely. Our exact
calculation shows that this maximal time 
scale $t_m(r)$ in the JC model is identical for both detection schemes mentioned above and does exhibit a 
single minimum at an optimal value $r_m^*$.
Finally, our results for arbitrary two level systems are generalized to non-Poissonian 
measurement protocols where the measurement events are still uncorrelated, but the waiting time between 
successive measurements are distributed via a general $p(\tau)$, not necessarily
exponential as in the Poissonian case.
We show that the short time behavior of
$F_r(t)\sim p(0)\, t^2$ is universal as long as $p(0)\ne 0$.

The rest of the paper is organised as follows: In Sec.~\ref{sec:SP}, we discuss the first 
detection probability and its 
cumulative distribution, which we call survival probability, when the system is 
subjected to unitary evolution and projective 
measurements at random times. This is done for a generic Hamiltonian and a generic random measurement protocol. In 
Sec.~\ref{sec:QMP}, assuming still a generic Hamiltonian, we discuss the Poissonian measurement protocol where a 
measurement is attempted at random continuous times with rate $r$ and we provide a general framework to compute the PDF 
of the first detection time. In Sec.~\ref{sec:GE}, we focus on a generic two-level system under Poissonian protocol 
where more explicit results can be obtained. We discuss two different detection schemes
: (i) when the state of interest differs from the initial state 
and (ii) when they are identical. For a generic two level Hamiltonian, we derive the exact Laplace transform of the PDF
of the first detection time and also discuss the exact asymptotic behavior of the
mean first detection time for small and large detection rate $r$. 
In Sec.~\ref{sec:JC_mfd}, we derive explicitly the PDF of the first detection time in 
the Jaynes-Cummings model for both detection schemes. 
In Sec. \ref{sec:general}. we provide a generalization of the non-Poissoinian measurement protocol
where the waiting time between independent measurements are drawn from a general $p(\tau)$. 
Sec.~\ref{sec:conc} contains a summary and an outlook. 
Some details of the calculations are relegated to the appendices.

\section{The first detection probability: general formalism}
\label{sec:SP}

Consider a general quantum system with Hamiltonian 
$H$, prepared in an initial state $|\psi(0)\rangle$. Without any external coupling to environment, the state
of the system evolves unitarily and at time $t$, it is given by
\begin{equation}
\label{eq:se}
| \psi(t) \rangle = e^{-i H t} | \psi(0) \rangle\, .
\end{equation}
Any attempt to measure or detect the system in some state results in an interruption of the
unitary evolution. Suppose, at time $t$, we make an attempt to detect if the system is in a specific 
state $|\psi_{\rm int}\rangle$, where the subscript `int' stands for `state of interest'.
Immediately after the measurement, the state of the system changes depending on the outcome
of the detection attempt. This new post measurement state can be characterized naturally in terms of the
pair of projection operators 
\begin{eqnarray}
\label{eq:P}
\hat{P}  &=& | \psi_{\rm int}\rangle   \langle  \psi_{\rm int} |  \\
\hat{Q} &=& I - \hat{P}  = I-  | \psi_{\rm int}\rangle   \langle  \psi_{\rm int} | . 
\label{eq:Q}
\end{eqnarray}
The operator $\hat{Q}$ projects a state to the complementary subspace of $| \psi_{\rm int}\rangle$.
If the outcome of the measurement is a success, then the state of the system immediately after the
measurement becomes $\hat{P}\, |\psi(t)\rangle$. In contrast, if the measurement is a failure, the
state becomes $\hat{Q}\, |\psi(t)\rangle$. This is precisely the projective measurement.
Following the measurement attempt, the system again evolves unitarily starting from $\hat{P}\, |\psi(t)\rangle$
or from $\hat{Q}\, |\psi(t)\rangle$, depending on the outcome of the first result, until another
measurement is attempted at a future time. Thus, successive projective measurements
are like successive `quantum resettings' of the state of the system -- resetting in this context simply
means hitting the state with a projection operator which in this case could be either $\hat{Q}$ or $\hat{P}$.

A question that arises naturally is: when is the first time $t_r$  
that the state of interest $|\psi_{\rm int}\rangle$
will be detected if one keeps making measurements? Here we use the subscript $r$ to denote
`resetting' in general.  
Obviously $t_r$ depends on the details of
the measurement protocol. In this paper, we will consider random measurements where attempts
at detecting the state of interest occurs at random times $T_1$, $T_2$, $T_3$ etc; In that
case, the first detection time $t_r$ also becomes a random variable with a probability
distribution function (PDF) denoted by $F_r(t)= {\rm Prob.}[t_r=t]$. 
Our main goal in this paper is to compute $F_r(t)$ for a class of
Hamiltonians $H$ and some specific random measurement protocols. 

This question concerning the first detection time in projective measurements has been addressed in a number of recent 
papers~\cite{DDD15,DDDS15,FKB17,TBK18,YZTB19,KEZ21,ZBK21,DG22}. Most of these papers addressed a stroboscopic 
measurement protocol where a measurement is attempted periodically with a fixed period $T$, with the exception of 
Refs.~\cite{KEZ21, ZBK21, DG22} where more general measurement protocols were discussed. Ref.~\cite{YZTB19, KEZ21} 
developed a renewal formalism for general random measurement protocols to compute the first detection probability 
directly and used it to study a quantum random walk model~\cite{YZTB19,DG22}. Below we present an alternative general 
formalism that is also adaptable to any measurement protocol. In our approach, it is more convenient to compute first 
the cumulative distribution of $F_r(t)$ (we call this survival probability below) and from it derive $F_r(t)$ by 
taking the derivative with respect to $t$, more in line with the approach used in Refs.~\cite{DDD15, DDDS15, DG22}.

As mentioned above,
instead of computing the PDF of $t_r$ 
directly, we find it more convenient to compute its cumulative distribution function (CDF) defined as
\begin{equation}
S_r(t)= {\rm Prob.}[t_r\ge t] = \int_t^{\infty} F_r(t')\, dt'\, .
\label{surv_ft.1}
\end{equation}
Thus, if we can compute $S_r(t)$, then the PDF of $t_r$ follows simply from the derivative
\begin{equation}
F_r(t)= - \frac{dS_r(t)}{dt}\, .
\label{fdt.1}
\end{equation}
The quantity $S_r(t)$ has a physical meaning: it is the probability that the state $|\psi_{\rm int}\rangle$
is {\em not detected} up to time $t$ under a given random measurement protocol. 
Since the process stops when the state of interest is detected for the first time,
we will call $S_r(t)$, i.e., the probability of not being detected up to $t$, as the
`survival probability' up to $t$.  
An observable of much practical importance is
the mean first detection time given by the first moment of $F_r(t)$, i.e., 
\begin{equation}
\bar{t}_{r}= \int_0^{\infty} t\, F_r(t)\, dt =\int_0^{\infty} S_r(t)\, dt\, ,
\label{moment.1}
\end{equation}
where we used Eq.~(\ref{fdt.1}) and performed integration by parts with the initial condition $S_r(0)=1$. 
Thus, knowing $S_r(t)$ allows us to compute easily the mean first detection time, as well as all
higher moments of $t_r$.  

The computation of $S_r(t)$ turns out to be simpler than
that of $F_r(t)$ since one can break up
$S_r(t)$ into a sum of terms representing the probability of $0$, $1$, $2$, $3$ etc. of 
failed detections up to $t$.
For example, if there is no measurement event up to $t$, then the state of the system at time $t$ is
\begin{equation}
|\psi_0(t)\rangle= e^{-i H t}\, |\psi(0)\rangle\, .
\label{n0.1}
\end{equation}
The subscript `$0$' in $|\psi_0(t)\rangle$ stands for no measurement.
Then the probability that the state $|\psi_{\rm int}\rangle$ is not detected up to $t$ is trivially
\begin{equation}
S_r(0,t) =\langle \psi_0(t)|\psi_0(t)\rangle 
= \langle \psi(0)|\psi(0)\rangle=1\, ,
\label{s0.1}
\end{equation}
where we assume that the initial state $|\psi(0)\rangle$ is normalized to unity.
Here $S_r(0,t)$ denotes the survival probability up to $t$, conditioned on having no measurement prior to $t$.
Now, suppose we make the first measurement at some time $\tau_1$, resulting in a failure. 
Then the state of the system
immediately after this first failed detection is
\begin{equation}
\label{eq:fail_detection}
|\psi_1(\tau_1) \rangle = \hat{Q} e^{-i H \tau_1} | \psi(0) \rangle\, .
\end{equation}
The subscript `$1$' denotes one measurement resulting in a failed detection.
At any subsequent time $t\ge \tau_1$, if there are no measurement attempts in the interval $[\tau_1, t-\tau_1]$, the state of the
system is given by
\begin{equation}
|\psi_1(t)\rangle= e^{-i H (t-\tau_1)}\hat{Q} e^{-i H \tau_1} | \psi(0) \rangle \, ,  \quad t\ge \tau_1 \, .
\label{st_f1.1}
\end{equation}
In this case, the survival probability $S_r(1,t|\tau_1)$ up to $t$, conditioned on
having one failed detection at $\tau_1<t$, is simply
\begin{equation}
\label{S1.1}
S_r(1,t|\tau_1)=     \langle \psi_{1}(t)  | \psi_{1}(t) \rangle  
=  \langle \psi(0)| e^{i H \tau_1}\, \hat{Q}^{\dagger}\, \hat{Q}\, e^{-i H \tau_1}|\psi(0)\rangle\, . 
\end{equation}
Using $\hat{Q}^{\dagger} \hat{Q}= \hat{Q}$ (since $\hat{Q}$ is a projection operator), this simplifies to
\begin{equation}
\label{eq:sp1}
S_r(1,t|\tau_1)\equiv S_r(1|\tau_1)=\langle \psi(0)| e^{i H \tau_1}\, \hat{Q}\, e^{-i H \tau_1}|\psi(0)\rangle\, .
\end{equation}
Note that $S_r(1,t|\tau_1)$ does not depend on $t$ explicitly, but
only on $\tau_1$. Hence, we denote it by $S_r(1|\tau_1)$ in Eq.~(\ref{eq:sp1}).
Finally, let us remark that till the time of the first measurement at $\tau_1$, the norm of the state
does not change from its initial value $\langle \psi(0)|\psi(0)\rangle=1$. 
But at any subsequent time $t$ after the 
first failed measurement at $\tau_1$, 
the norm $S_r(1,t|\tau_1)$ is less than unity.
Thus every time a measurement takes place it reduces the norm of the state.

\begin{figure}
\centering
\includegraphics[width=0.7\linewidth]{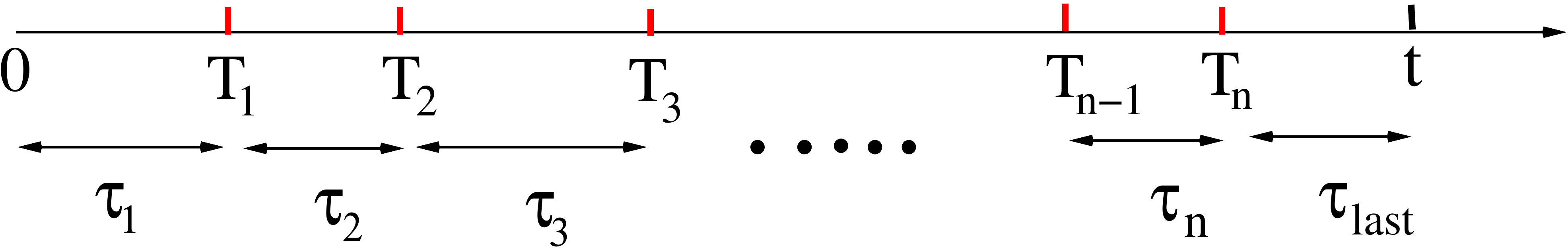}
\caption{A schematic representation of a configuration
of the random measurement protocol with $n$ measurements taking
place till time $t$ at random epochs $T_1$, $T_2$, $T_3$, $\ldots$, $T_n$. The interval between 
successive measurements are denoted by $\tau_k= T_k-T_{k-1}$, with $k=1,2,\ldots, n$. The last
interval $\tau_{\rm last}=t-T_n= t-\sum_{k=1}^n \tau_k$ is `measurement free'.}
\label{fig:schem}
\end{figure}

Now, consider the case when there are two failed measurements, respectively at $T_1=\tau_1$
and $T_2= \tau_1+\tau_2$, prior to $t$. The state of the system at time $t\ge \tau_1+\tau_2$ is then given by
\begin{equation}
|\psi_2(t)\rangle= e^{-i H \left(t-(\tau_1+\tau_2)\right)}\, \hat{Q}\, e^{-i H \tau_2}  \hat{Q}\,
e^{-i H \tau_1} | \psi(0) \rangle \, .
\label{st_f2.1}
\end{equation}
The subscript $2$ in $|\psi_2(t)\rangle$ indicates $2$ measurements both resulting in failed detections.
In this case, the survival probability $S_r(2,t|\tau_1, \tau_2)$ up to $t$, conditioned on
having two failed detections respectively at $\tau_1$ and $\tau_1+\tau_2$, is simply
\begin{equation}
\label{S2.1}
S_r(2,t|\tau_1,\tau_2)=     \langle \psi_{2}(t)  | \psi_{2}(t) \rangle
=  \langle \psi(0)| e^{i H \tau_1}\, \hat{Q}^{\dagger} e^{i H \tau_2} \hat{Q}^{\dagger}\, \hat{Q}\, 
e^{-i H \tau_2} \hat{Q} e^{-i H \tau_1}|\psi(0)\rangle\, .  
\end{equation}
Using $\hat{Q}^{\dagger} \hat{Q}= \hat{Q}$, this reduces to
\begin{equation}
\label{eq:sp2}
S_r(2,t|\tau_1,\tau_2)\equiv  S_r(2|\tau_1,\tau_2) =
\langle \psi(0)| e^{i H \tau_1}\, \hat{Q}^{\dagger} e^{i H \tau_2} \hat{Q}\, 
e^{-i H \tau_2} \hat{Q} e^{-i H \tau_1}|\psi(0)\rangle\, . 
\end{equation}
As in the case $n=1$, we note that $S_r(2,t|\tau_1,\tau_2)$ does not depend on $t$ explicitly,
hence we denote it by $S_r(2|\tau_1,\tau_2)$.

We can then easily generalise to the case when there are $n$ measurements prior to $t$
all resulting in failed detections of
the state of interest $|\psi_{\rm int}\rangle$, 
taking place at
epochs $\{T_1 =\tau_1$, $T_2=\tau_1+\tau_2$, $T_3=\tau_1+\tau_2+\tau_3$, $\cdots$, $T_n=\sum_{k=1}^n \tau_k \}$.
These measurement epochs 
are separated by time intervals $\{\tau_1,\tau_2,\tau_3,\ldots, \tau_n\}$ [see Fig. (\ref{fig:schem})]. 
If there are $n>0$
such failed attempts prior to $t$,
then the state of the system at time $t>T_n$, conditioned on having $n>0$ failed
attempts prior to $t$, is given by
\begin{equation}
|\psi_n(t)\rangle= e^{-i H \tau_{\rm last}}\,\prod_{k=1}^n \hat{Q}\, e^{-i H \tau_k}\, |\psi(0)\rangle\, ,
\label{st_fn.1}
\end{equation}
where $\tau_{\rm last}= t-T_n$ denotes
the length of the `measurement free' last interval before $t$ [see Fig.~(\ref{fig:schem})].
The survival probability up to $t$, conditioned on having $n>0$ such failed attempts before $t$
occurring at epochs separated by intervals $\{\tau_i\}\equiv \{\tau_1,\tau_2,\ldots, \tau_n\}$, is
then given by the norm $\langle \psi_{n}(t)  | \psi_{n}(t) \rangle$.
From Eq.~(\ref{st_fn.1}), it is clear that upon taking the norm, the contribution
from the last interval $ e^{-i H \tau_{\rm last}}$ disappears since it
represents the free unitary
evolution following the last measurement till $t$ as shown in Fig.~(\ref{fig:schem}).
Consequently the resulting survival probability $S_r(n,t|\{\tau_i\})$ becomes
independent of the last inteval $\tau_{\rm last}$ and is given by 
\begin{eqnarray}
\label{sn.1}
S_r(n,t|\{\tau_i\}) &=&  \langle \psi_{n}(t)  | \psi_{n}(t) \nonumber \\
&=& \langle \psi(0)|\left[ \prod_{k=1}^n e^{i H \tau_k}\,\hat{Q}^{\dagger}\right]\,  
\left[\, \prod_{m=1}^{n}\hat{Q}\, e^{-i H \tau_{n-m+1}}\right]|\psi(0)\rangle \nonumber \\
&\equiv& S_r(n|\{\tau_i\})\, .
\end{eqnarray}
Recall that $\hat{Q}^{\dagger}=\hat{Q}$ given in Eq.~(\ref{eq:Q}).
As in the case of $n=1$ and $n=2$, we note that $S_r(n,t|\{\tau_i\})$ in Eq.~(\ref{sn.1})
does not depend on $t$ explicitly, but only on the interval
lengths $\{\tau_i\}$ that occur till the $n$-th measurement preceding $t$.
Hence, we denote it simply by $S_r(n|\{\tau_i\})$ in Eq.~(\ref{sn.1}).
Note that the case $n=0$ is a bit special since there is no measurement, and consequently no failed attempt
and $S_r(0,t)=1$ as in Eq.~(\ref{s0.1}).
However, for convenience, we can still use the general formula in Eq.~(\ref{sn.1})
even for $n=0$ provided the intervening operator between $\langle \psi(0)|$ and
$|\psi(0)\rangle$ is identified with the identity operator $I$. Thus, with this convention,
$S_r(n,t|\{\tau_i\})$ in Eq.~(\ref{sn.1}) also includes the $n=0$ case. 

Any measurement configuration showing `no detection' up to $t$ is fully characterized by the 
number of measurements $n$ that take place up to time $t$, as well as by the random time intervals
$\{\tau_1,\tau_2,\ldots\,,\tau_n\}$ between two successive measurements (all measurement events
resulting in failed detections) in
a given configuration, as shown in Fig.~(\ref{fig:schem}). Clearly, given $t$,
both $n$ and the intervals $\{\tau_i\}=\{\tau_1,\tau_2,\ldots, \tau_n\}$ are random variables
whose statistics depend on the precise nature of the random measurement protocol. Generically, let
$P(n, \{\tau_i\}|t)$ denote the joint distribution of $n$ and
the set $\{\tau_i\}$ , given a fixed time $t$.
In that case, considering all possible random number of measurements and the associated
intervals between successive measurements till $t$,
the total survival probability up to time $t$ can be expressed by the compact formula
\begin{equation}
S_r(t)=  
\sum_{n=0}^{\infty}\int_0^{\infty}d\tau_1\dots\int_0^{\infty}d\tau_n\, 
P(n, \{\tau_i\}|t)\, S_r(n|\{\tau_i\})\, ,
\label{surv.1}
\end{equation} 
where $S_r(n|\{\tau_i\})$ is given in Eq.~(\ref{sn.1}).
Note that in this formula, quantum operators and states appear only in
$S_r(n|\{\tau_i\})$, but not in $P(n, \{\tau_i\}|t)$ which
is a pure classical probability that characterizes the random
measurement protocol.
Thus the total survival probability $S_r(t)$ up to $t$ involves a
convolution of quantum and classical dynamics, the later being induced
by the stochastic projective measurements.
 
Even though the formula in Eq.~(\ref{surv.1}) is exact for a generic Hamiltonian and a generic measurement
protocol, it is somewhat formal. Our goal in
the next two sections is to show that for Poissonian measurement protocol and
for a generic two-level Hamiltonian, the formal expression in Eq.~(\ref{surv.1})
simplifies enormously and consequently, 
the survival probability $S_r(t)$,
or equivalently the first detection probability $F_r(t)=-dS_r(t)/dt$, can be computed explicitly.

\section{Poissonian Measurement Protocol}
\label{sec:QMP}

In this section we focus on a specific random measurement protocol known as the Poissonian protocol, for
which the joint probability density $P(n, \{\tau_i\}|t)$ can be written in a nice explicit form.
Consequently, the result in Eq.~(\ref{surv.1}) simplifies considerably and takes a compact
form in the Laplace space conjugate to $t$.

In the Poissonian protocol, the measurements take place stochastically in continuous time with a constant rate $r$.
More precisely, in a small time $dt$, a measurement takes place with probability $r\,dt$ and with
the complementary probability $(1-r\,dt)$ no measurement occurs. This measurement process is
Markovian since the probability of a measurement at the instant $t$ is independent of the
history of the measurement process till $t$. The probability
that an interval $[0,t]$ is `measurement free' is then simply $e^{-r t}$. Furthermore, the intervals
between successive measurements are statistically independent and the PDF of each interval $\tau_k$ is
simply exponential, i.e., $p(\tau)={\rm Prob.}[\tau_k=\tau]= r\, e^{-r\, \tau}$. This follows from
the fact that after the $(k-1)$-th measurement, one should have a measurement free interval
of length $\tau$ (the probability of this event is $e^{-r \tau}$), followed by
a measurement event at the end of this interval which occurs with probability $r\, d\tau$.
Consequently, the PDF is $p(\tau)={\rm Prob.}[\tau_k=\tau]= r\, e^{-r\, \tau}$, which is normalized to unity.
The Poissonian protocol appears quite naturally in classical stochastic resetting problems
where $r$ denotes the rate of resetting~\cite{EM11,EM12,EMS20}. It also appears quite
naturally in the run-and-tumble dynamics of active particles where $r$ denotes the 
tumbling rate~\cite{MLMS20,MLMS20.2}.

We consider a `no detection' measurement configuration up to a fixed $t$ 
that is fully characterized by $n$ (the number of measurements before $t$),
the interval lengths $\{\tau_1,\tau_2,\tau_3,\ldots,\tau_n\}$ between
measurements, as depicted in  Fig.~(\ref{fig:schem}). Given $t$ and
the set $\{\tau_1,\tau_2,\tau_3,\ldots,\tau_n\}$, the length of
the last interval gets fixed $\tau_{\rm last}= t- \sum_{k=1}^n \tau_k$.
Using the statistical independence of successive intervals, the joint
probability density $P(n, \{\tau_i\}|t)$, given a fixed $t$, can then be
written as the simple product~\cite{MLMS20,MLMS20.2}
\begin{equation}
P(n, \{\tau_i\}|t) =  \left[\prod_{k=1}^n r\, e^{-r\, \tau_k}\right]\, 
e^{-r\, \left(t- \sum_{k=1}^n \tau_k\right)}\, .
\label{jpdf_product}
\end{equation}
Note that, each of
the $\tau_k$'s is distributed independently via the PDF $p(\tau)=r\, e^{-r \,\tau}$, but
the last interval $\tau_{\rm last}=t-\sum_{k=1}^n \tau_k$ is a bit apart. The statistical weight associated to it
is $e^{-r\, \tau_{\rm last}}$ without the additional multiplicative factor $r$ unlike the
intervals preceding it -- this fact has been used and exploited in
many recent contexts~\cite{MLMS20,MLMS20.2,SM22,SMS23}. This is due to the fact that the last interval 
is measurement free and
the probability for that is just $e^{-r \,\tau_{\rm last}}$. Thus the Poissonian protocol
can be viewed as a classical continuous time renewal process with exponentially distributed
intervals~\cite{Feller}

It is useful to check that the joint distribution in Eq.~(\ref{jpdf_product}) is appropriately
normalized to unity when one integrates over $\tau_k$'s and sums over $n$. To check this,
it is first convenient to express the joint PDF in Eq.~(\ref{jpdf_product}) as an integral
representation
\begin{equation}
P(n, \{\tau_i\}|t) =  \int_0^{\infty} d\tau_{\rm last} 
\left[\prod_{k=1}^n r\, e^{-r\, \tau_k}\right]\, e^{-r\, \tau_{\rm last}}\,
\delta\left(\tau_1+\tau_2+\ldots +\tau_n+ \tau_{\rm last}-t\right)\, .
\label{jpdf.1}
\end{equation}
The presence of the delta function naturally suggests to take the Laplace transform with respect to $t$.
This gives, upon integrating over $\tau_{\rm last}$ trivially,
\begin{equation}
\int_0^{\infty} P(n, \{\tau_i\}|t)\, e^{-s\, t}\, dt =
\frac{1}{r+s}\, \left[\prod_{k=1}^n r\, e^{-(r+s)\, \tau_k}\right]\, .
\label{jpdf_lap.1}
\end{equation}
Now integrating over the set $\{\tau_i\}$ over $\tau_i\in [0,\infty]$, we get the Laplace transform of the marginal
distribution $P(n|t)$ denoting the distribution of the number of measurements up to $t$,
\begin{equation}
\int_0^{\infty} P(n|t)\, e^{-s\, t}\, dt= \frac{r^n}{(r+s)^{n+1}}\, .
\label{pns.1}
\end{equation}
Inverting this Laplace transform gives the Poisson distribution
\begin{equation}
P(n|t)= e^{-r\, t}\, \frac{(r\,t)^n}{n!}\, \quad n=0,1,2,\ldots\, ,
\label{poisson.1}
\end{equation}
which justifies the name Poissonian protocol. Evidently $\sum_{n=0}^\infty P(n|t)=1$, thus verifying
that the joint PDF in Eq.~(\ref{jpdf_product}) is correctly normalized to unity.

We now substitute the result Eq.~(\ref{jpdf.1}) in the general formula for the survival probability
in Eq. (\ref{surv.1}). This gives a rather long expression
\begin{equation}
S_r(t) =   \sum_{n=0}^{\infty} \int_0^{\infty} d\tau_{\rm last}\, e^{-r\, \tau_{\rm last}}\,
\left[ \prod_{k=1}^n \int_0^{\infty} d\tau_k\, r\, e^{-r\, \tau_k}\right]\, 
S_r(n|\{\tau_i\})\, \delta\left(\tau_1+\dots+\tau_n+\tau_{\rm last}-t\right) \, .  
\label{surv.2}
\end{equation}
Taking Laplace transform of Eq.~(\ref{surv.2}) with respect to $t$ and performing
the integral over $\tau_{\rm last}$ explicitly, one arrives at a slightly more compact expression
\begin{equation}
\tilde{S}_r(s) =   \int_0^{\infty} dt\, e^{-s\, t}\, S_r(t)  
= \frac{1}{r+s}\, \sum_{n=0}^{\infty} 
\left[ \prod_{k=1}^n \int_0^{\infty} d\tau_k\, r\, e^{-(r+s)\, \tau_k}\, 
S_r(n|\{\tau_i\})\right]\, , 
\label{surv_lap.0}
\end{equation}
where $S_r(n|\{\tau_i\})$ is given in Eq.~(\ref{sn.1}). 
Note that the first term in the summation corresponding to $n=0$ is simply $1/(r+s)$. Hence it
is useful to separate out the $n=0$ term from the sum and write it as
\begin{equation} 
\tilde{S}_r(s) =  \int_0^{\infty} dt\, e^{-s\, t}\, S_r(t) 
= \frac{1}{r+s} + \frac{1}{r+s}\, \sum_{n=1}^{\infty}
\left[ \prod_{k=1}^n \int_0^{\infty} d\tau_k\, r\, e^{-(r+s)\, \tau_k}\,
S_r(n|\{\tau_i\})\right]\, .  
\label{surv_lap.1}
\end{equation}
This result 
in Eq.~(\ref{surv_lap.1}) for the Poissonian protocol is valid for generic Hamiltonian $H$.
However, Eq.~(\ref{surv_lap.1}) is still a bit formal because for a generic Hamiltonian $H$,
it is not easy to compute explicitly $S_r(n|\{\tau_i\})$ in Eq.~(\ref{sn.1}). In the next
section, we show that for a two-level Hamiltonian $H$, one can compute $S_r(n|\{\tau_i\})$
explicitly. This leads to a simpler expression of the main result in Eq.~(\ref{surv_lap.1}).

\section{Arbitrary two-level system with Poissonian measurement protocol}
 \label{sec:GE}

In the previous section, we derived the Laplace transform of the survival probability
$S_r(t)$ up to time $t$ [see Eq.~(\ref{surv_lap.1})] for a system with a generic Hamiltonian $H$, subjected to
Poissionian measurement protocol with rate $r$. As mentioned above,
this formula [Eq.~(\ref{surv_lap.1})] still contains the quantities $S_r(n|\{\tau_i\})$ 
which are hard to compute in practice
for a general quantum Hamiltonian $H$ with a large Hilbert space.
In this section, we show that they can be computed in closed form for
a general two-level system where the dimension of the Hilbert space is just two.
The reason behind this is that the projection operator $\hat{Q}$ involves only
one state which makes the computation much simpler, as we illustrate below. 

To be precise, we consider a general two-level Hamiltonian with two possible states
in the Hilbert space, denoted respectively by $|\psi_+\rangle$ and $| \psi_-\rangle$.
Without any loss of generality, we assume that the system is prepared at $t=0$ in
the $+$ state, i.e., $|\psi(0)\rangle= |\psi_+\rangle$. The complementary case when
the system is initially in state $|\psi_{-}\rangle$ can be treated in a similar way.
We will now consider two different schemes: 

\begin{itemize}

\item Scheme $1$: when the state of interest $|\psi_{\rm int}\rangle$
differs from the initial state $|\psi_+\rangle$, i.e., when $|\psi_{\rm int}\rangle= 
|\psi_{-}\rangle$. 

\item Scheme $2$: when the state of interest coincides with the initial state, i.e.,
when $|\psi_{\rm int}\rangle= |\psi_+\rangle$.

\end{itemize}
These two schemes are summarized in Table.~\ref{tab_table1}. We remark that
the second scheme has been studied recently for general two level systems subject 
to random measurement protocols using a different method~\cite{KEZ21,ZBK21}--where it is referred to as the `return problem'.
Here we will show that our method allows to derive explicit results for both schemes under
the same footing and also to compare the respective outcomes. We will show that the two schemes
lead to very different results for the distribution of the first detection time and
its moments. In particular, the short time behavior of the PDF of the first detection time
as well as the mean detection time 
have drastically different behaviors in the two schemes.

\begin{table}
\centering
\begin{tabular}{|c|c||c||c||c|}
\hline
 & Initial State  $|\psi (0) \rangle$ &  State of interest  
$|\psi_{\rm int} \rangle$ & $|\psi_c \rangle$   & $f(\tau), g(\tau)$ \\ 
  \hline
1 &  $| \psi_+\rangle$ & $| \psi_-\rangle$ & $| \psi_+\rangle$  & $f(\tau)=g(\tau)$ \\ 
  \hline
 2 &  $| \psi_+\rangle$ & $| \psi_+\rangle $ & $| \psi_-\rangle$  & $f(\tau)+g(\tau)=1$ \\ 
 \hline
 \end{tabular}
 \caption{The table demonstrates the detection schemes for a general two level system 
with two states 
$| \psi_+\rangle$ and $| \psi_-\rangle$ in its Hilbert space. The system
is prepared initially in the $+$ state, i.e., 
$|\psi(0)\rangle=|\psi_+\rangle$.
There are two schemes depending on the state of interest $|\psi_{\rm int}\rangle$.  
The scheme $1$ corresponds to a situation when the initial state  
$| \psi(0) \rangle =|\psi_+\rangle $ is different from the state of interest 
$|\psi_{\rm int} \rangle=|\psi_{-}\rangle$ . The scheme $2$ corresponds to the reverse situation 
when the initial state $|\psi(0)\rangle=|\psi_+\rangle $ is the
same as the state of interest, i.e., $|\psi_{\rm int}\rangle=|\psi_{+}\rangle$.
The coloumn $| \psi_c\rangle$ represents the state complementary to the state of 
interest (i.e., the other state of the two level system). The last 
column $f(\tau)$ and $g(\tau)$ are given in Eq.~\eqref{eq:fdef} and Eq.~\eqref{eq:gdef} respectively. }
  \label{tab_table1}
\end{table}

We start by showing how the formula for $S_r(n|\{\tau_i\})$ in Eq.~(\ref{sn.1})
simplifies for a two-level system such that
the Laplace transform $\tilde{S}_r (s)$ in Eq.~(\ref{surv_lap.1})
can be expressed in terms of only
two matrix elements. The key to the simplication for a two-level system
lies in the fact that the projection operator 
$\hat{Q}= I-|\psi_{\rm int}\rangle\langle \psi_{\rm int}|= |\psi_c\rangle\langle \psi_c|$
involves a single state $|\psi_c\rangle$ which is complementary to $|\psi_{\rm int}\rangle$
in either scheme. The subscript `c' in $|\psi_c\rangle$ stands for `complementary'.
Consequently $S_r(n|\{\tau_i\})$
etc. can be expressed very simply. For example, for $n=1$, using $|\psi(0)\rangle=|\psi_+\rangle$
and $\hat{Q}=|\psi_c\rangle\langle \psi_c|$ in
Eq. (\ref{eq:sp1}), we get
\begin{equation}
\label{eq:s1ta_eg}
S_r(1|\tau_1)  =   \langle   \psi_+  |  e^{i H \tau_1}  
|\psi_c \rangle \langle \psi_c 
| e^{-i H \tau_1} |\psi_+ \rangle = 
\big|\langle   \psi_+ |  e^{i H \tau_1} | \psi_c \rangle\big|^2 \, . 
\end{equation}
Similarly, from Eq.~\eqref{eq:sp2}
\begin{equation}
\label{eq:s1t_eg}
S_r(2|\tau_1,\tau_2)  
= \big| \langle   \psi_+  |  e^{i H \tau_1}  
|\psi_c \rangle \big|^2 \big| \langle\psi_c | e^{i H \tau_2}  
| \psi_c \rangle \big|^2. 
\end{equation}
Likewise, for general $n$ in Eq.~(\ref{sn.1}), we have 
\begin{equation}
\label{eq:snt_eg}
S_r(n|\{\tau_i\})= \big| \langle   \psi_+  |  e^{i H \tau_1}
|\psi_c \rangle \big|^2\, \prod_{k=2}^n \big| \langle\psi_c | e^{i H \tau_k}
| \psi_c \rangle \big|^2 \, ,
\end{equation}
where $\{\tau_i\}\equiv \{\tau_1,\tau_2,\ldots, \tau_n\}$.
For convenience, we define a pair of functions
\begin{eqnarray}
\label{eq:fdef}
f (\tau) &\equiv&   \big| \langle   \psi_+  |  e^{i H \tau}  | \psi_c \rangle \big|^2  \\
g (\tau) &\equiv&    \big|   \langle \psi_c  \mid e^{i H \tau} | \psi_c  \rangle   \big|^2
\label{eq:gdef}
\end{eqnarray}
and their Laplace transforms
\begin{eqnarray}
\int_0^{\infty} d\tau\, f (\tau)\, e^{-s\tau} &= & \tilde{f}(s),\quad \\
\int_0^{\infty} d\tau\, g (\tau)\, e^{-s\tau} &= & \tilde{g}(s)\, . 
\label{eq:fg_laplace}
\end{eqnarray}

In terms of the two functions $g(\tau)$ and $f(\tau)$ defined above, Eq.~(\ref{eq:snt_eg}) reads 
\begin{equation}
S_r(n|\{\tau_i\})= f(\tau_1)\, \prod_{k=2}^n g(\tau_k) \, .
\label{eq:sn}
\end{equation}
We next substitute the result in our general formula in Eq.~(\ref{surv_lap.1}).
In terms of the Laplace transform in Eq.~(\ref{eq:fg_laplace}) we get
\begin{equation}
\tilde{S}_r(s)=  \frac{1}{r+s}\left[ 
1+ r\, \tilde{f}(r+s)\, \sum_{n=1}^{\infty} \left[r\, \tilde{g}(r+s)\right]^{n-1}\right] 
= \frac{1}{r+s}\left[ 1+ \frac{r\, \tilde{f}(r+s)}{1- r\, \tilde{g}(r+s)}\right]\, .  
\label{eq:s_tilde_fg_laplace}
\end{equation}
This exact Laplace transform requires the knowledge of just two function $f(\tau)$ and $g(\tau)$
defined respectively in Eqs.~(\ref{eq:fdef}) and (\ref{eq:gdef}). Thus, essentially, one
needs to just compute two matrix elements $\langle \psi_+|e^{i H \tau}|\psi_c\rangle$ and
$\langle \psi_c|e^{i H \tau}|\psi_c\rangle$. These matrix elements depend, of course, on
the details of the two-level Hamiltonian $H$.
But even without computing these matrix elements explicitly, it is possible to extract
some general properties of the PDF
$F_r(t)$ for a generic two-level Hamiltonian, as demonstrated below.

\vspace{0.4cm}

\noindent {\em Probability distribution of the first detection time.}
We recall from Eq.~(\ref{fdt.1}) that $F_r(t)=-dS_r(t)/dt$. Taking
Laplace transform with respect to $t$ on both sides and using $S_r(0)=1$, it follows that
\begin{equation}
\tilde{F}_r(s) = \int_0^{\infty} F_r(t)\, e^{-s\, t}\, dt= 1- s \tilde{S}_r(s) 
= 1 - \frac{s}{r+s}\bigg[\frac{1+r\big(\tilde{f}(r+s) - \tilde{g}(r+s)  \big)}{1-r\tilde{g}(r+s)} \bigg],
\label{eq:tildeF}
\end{equation}
where we used Eq.~\eqref{eq:s_tilde_fg_laplace}. This Laplace transform can be
formally inverted as  
\begin{equation}
F_r(t) = \int_{\Gamma} \frac{ds}{2\pi i }\, e^{st}\, \tilde{F}_r(s).
\label{eq_inv_lt_def}
\end{equation}
where $\Gamma$ denotes a Bromwich contour in the complex $s$ space. While inverting this
Laplace transform explicitly is hard and it depends on the specific form of $\tilde{g}(s)$, we will
see below that some asymptotic behaviors, such as the small $t$ behavior of $F_r(t)$,
can be extracted explicitly.

\vspace{0.4cm}

\textit{Mean first detection time.}
The mean first detection time in Eq.~(\ref{moment.1}), upon setting $s=0$ in Eq.~(\ref{eq:s_tilde_fg_laplace}), 
is given by 
\begin{equation}
\bar{t}_r = \int_0^{\infty} t\, F_r(t)\, dt = \tilde{S}_r (0) =
 \frac{1}{r}  \bigg[ 1+ \frac{ r  \tilde{f}(r)}{1-r \tilde{g}(r)} \bigg]. 
\label{eq:ftd_main}
\end{equation}

Below, we analyse $F_r(t)$ and its first moment $\bar{t}_{r}$ for the two schemes separately. 

\subsection{First scheme}
\label{eq:gen_first_scheme}

We start with the first scheme mentioned in Table~\ref{tab_table1},
where $|\psi_{\rm int}\rangle= |\psi_{-}\rangle$. Its
complementary state is $|\psi_c\rangle= |\psi_+\rangle$. In this case,
it follows from Eqs.~(\ref{eq:fdef}) and (\ref{eq:gdef}) that
\begin{equation}
f(\tau)=g(\tau)=\big| \langle   \psi_+  |  e^{i H \tau}  | \psi_+ \rangle \big|^2 \, .
\label{fg.S1}
\end{equation} 
Hence, from Eq.~(\ref{eq:fg_laplace}), we have
$\tilde{f}(s) = \tilde{g}(s)$. Consequently, Eq. (\ref{eq:s_tilde_fg_laplace})
simplifies to 
\begin{equation}
\tilde{S}_r (s) =  \frac{1}{(r+s)}  \frac{1}{[1-r \tilde{g}(r+s)]}. 
\label{eq:s_tilde_fg_laplace_simp}
\end{equation}
Using Eq.~\eqref{eq:tildeF}, we get
\begin{equation}
\label{eq:ftilde_P1_gen}
\tilde{F}_r(s)=1- s\, \tilde{S}_r(s) 
= 1 - \frac{s}{r+s}\bigg[\frac{1}{1-r\tilde{g}(r+s)} \bigg]\, .
\end{equation}
Thus we have only one function $\tilde{g}(s)$ that characterizes
fully the probability distribution of the first detection time.

\vspace{0.4cm}

\noindent {\em Asymptotic behavior of $F_r(t)$ for small and large $t$.}
Given a generic $\tilde{g}(s)$, one can then easily deduce the small $t$ behavior
of $F_r(t)$ by analysing the large $s$ behavior of $\tilde{F}_r(s)$ in 
Eq.~(\ref{eq:ftilde_P1_gen}). For this we need to know how $\tilde{g}(s)$ behaves
for large $s$. From Eq.~(\ref{fg.S1}) we have, for scheme $1$,
\begin{equation}
\tilde{g}(s)= \int_0^{\infty} d\tau\, e^{-s\, \tau}\, \Big|\langle 
\psi_+ | e^{i H \tau}|\psi_+\rangle
\Big|^2\, .
\label{def_gr.1}
\end{equation}
The large $s$ behavior of $\tilde{g}(s)$ is governed by the short time behavior
of $g(\tau)$ in Eq.~(\ref{fg.S1}). To extract this large $s$ behavior
from Eq.~(\ref{def_gr.1}), it is useful to first make a change of variable 
$s\,\tau=\tilde{\tau}$ and rewrite the integral in Eq.~(\ref{def_gr.1}) as
\begin{equation}
\tilde{g}(s)= \frac{1}{s}\, \int_0^{\infty} d\tilde{\tau}\, e^{-\tilde{\tau}}\,  \Big|\langle
\psi_+| e^{i H\, \frac{\tilde{\tau}}{s}}|\psi_+\rangle
\Big|^2\, .
\label{def_gr.2}
\end{equation}
Next we expand $e^{i H \frac{\tilde{\tau}}{s}}$ in a power series of $1/s$ for large $s$. Keeping terms
up to $O(s^{-2})$ we get
\begin{equation}
\tilde{g}(s)= \frac{1}{s}\, \int_0^{\infty} d\tilde{\tau}\, 
e^{-\tilde{\tau}}\, \Big|\langle
\psi_+\Big| 1+ i H \frac{\tilde{\tau}}{s} - H^2\,
\frac{\tilde{\tau}^2}{2 s^2}+O\left(\frac{\tilde{\tau}^3}{s^{3}}\right)\, 
\Big|\psi_+\rangle \Big|^2.
\label{def_gr.3}
\end{equation}
This gives 
\begin{equation}
s\, \tilde{g}(s)= 1 - \frac{2}{s^2}\, 
\left[ \langle \psi_+| H^2|\psi_+\rangle-  
\Big|\langle \psi_+| H|\psi_+\rangle\Big|^2\right] + O(s^{-3})\, .
\label{def_gr.4}
\end{equation}
Therefore, we get 
\begin{equation}
\tilde{g}(s) = \frac{1}{s} - \frac{2 \sigma^2}{s^3}  + \dots
\label{eq:gen_g_large_s}
\end{equation}
where $\sigma^2$ is defined as 
\begin{equation}
 \sigma^2= \left[ \langle \psi_+|H^2|\psi_+\rangle-  
\Big|\langle \psi_+|H|\psi_+\rangle\Big|^2\right] \, . 
\label{eq:sigma_p1}
\end{equation}
Note that Eq.~(\ref{eq:gen_g_large_s}) implies 
\begin{equation}
g(\tau)= 1- \sigma^2\, \tau^2 +\ldots\quad {\rm as}\quad \tau\to 0\, .
\label{g_smallt}
\end{equation}

The quantity $\sigma^2$ has a nice interpretation:
it is the expected variance of the energy in the state $|\psi_+\rangle$. 
Its expression in Eq.~(\ref{eq:sigma_p1}) can be further simplified
by noting that
\begin{eqnarray}
\langle \psi_+| H^2|\psi_+\rangle &=&  \langle \psi_+\Big|H \left(\Big|\psi_+\rangle\langle 
\psi_+\Big|
+\Big|\psi_{\rm -}\rangle\langle \psi_{-}\Big|\right)\, H\Big|\psi_+\rangle \nonumber \\
&=&
\Big|\langle
\psi_+| H|\psi_+\rangle\Big|^2 + \Big|\langle \psi_+|H|\psi_{-}\rangle\Big|^2 \, .
\label{var.1}
\end{eqnarray}
Substituting Eq.~\eqref{var.1} into Eq.~(\ref{eq:sigma_p1}) gives
\begin{equation}
\sigma^2= \Big|\langle \psi_+|H|\psi_{-}\rangle\Big|^2 \, .
\label{sigma2.1}
\end{equation}
Substituting Eq.~(\ref{eq:gen_g_large_s})
into Eq.~(\ref{eq:ftilde_P1_gen}) gives, for large $s$,
\begin{equation}
\tilde{F}_r(s) \approx \frac{2\, r\, \sigma^2}{s^3}\, .
\label{Frs_large.1}
\end{equation}
Consequently, for small $t$, we obtain the following result for scheme $1$, 
valid for arbitrary two-level system,
\begin{equation}
F_r(t) \approx r\, \sigma^2\, t^2\, , \quad {\rm as}\quad t\to 0\, ,
\label{Frt_small.1}
\end{equation} 
with $\sigma^2$ given in Eq.~(\ref{sigma2.1}). This early time $\sim t^2$ behavior of $F_r(t)$
emerges from purely quantum dynamics and is universal, i.e., independent
of specific measurement protocols [see section~\eqref{sec:general}].

The large $t$ asymptotic behavior of $F_r(t)$ is more difficult to extract explicitly from its Laplace transform
in Eq.~(\ref{eq:ftilde_P1_gen}). However one can easily guess the general
form of the asymptotic decay at late times. Depending on $\tilde{g}(s)$, the Laplace 
transform $\tilde{F}_r(s)$ in Eq.~(\ref{eq:ftilde_P1_gen}) has typically several simple poles in
the complex $s$-plane, all with negative real parts. From Eq.~(\ref{eq:ftilde_P1_gen}),
these are given by the roots of $1- r\, \tilde{g}(r+s)=0$. The root whose
negative real part is closest to the origin will control the leading late
time exponential decay of $F_r(t)$, i.e., we expect   
$F_r(t) \sim e^{-t/t_m(r)}$ where $1/t_m(r)$ must satisfy $1- r\, \tilde{g}(r-1/t_m(r))=0$.
In the next section, we demonstrate this computation explicitly in the Jaynes-Cummings model.

\vspace{0.4cm}

\noindent {\em Mean first detection time.}
For the mean first detection time in Eq.~(\ref{eq:ftd_main}), one can however extract more detailed
information for a generic two level system.
For scheme 1, using $\tilde{f}(r)=\tilde{g}(r)$, one gets  
\begin{equation}
\bar{t}_r = \frac{1}{r}  \frac{1}{\left[1-r\, \tilde{g}(r)\right]} \, . 
\label{eq:ftd}
\end{equation}
One can then compute the asymptotic behaviors of $\bar{t}_{r}$ for small
and large $r$ for any generic two-level system in terms of the small $r$
and large $r$ behaviors of $\tilde{g}(r)$. 
Using Eq.~\eqref{eq:gen_g_large_s} in 
Eq.~(\ref{eq:ftd}), one immediately finds, for large $r$,
\begin{equation}
\bar{t}_{r}\approx \frac{r}{2\, \Big|\langle \psi_+|H|\psi_{-}
\rangle\Big|^2 }\, , \quad {\rm as}\quad r\to \infty\, .
\label{tr_final.1}
\end{equation}
The fact that the mean detection time diverges as $r\to \infty$ in scheme $1$ can be
easily understood as follows. When $r\to \infty$, the measurement takes place extremely frequently.
Consequently, the state of the system essentially does not evolve from its initial state $|\psi_+\rangle$.
This is indeed the Zeno limit. So, in scheme $1$, the state of interest 
$|\psi_{\rm int}\rangle=|\psi_{-}\rangle$ will never be detected in the large $r$ limit.
Naturally, the mean detection time also diverges as $r\to \infty$ in scheme $1$.

To compute the small $r$ behavior of $\bar{t}_{r}$ we use a different strategy.
We again start from the exact expression of $\tilde{g}(r)$ in Eq.~(\ref{def_gr.1}).
We first expand $|\psi_+\rangle$ in the eigenbasis of $H$ and write
\begin{equation}
|\psi_+\rangle= \sum_{E} a_E\, |E\rangle\, ,
\label{expand.1}
\end{equation}
where $|E\rangle$ denotes an eigenvector of $H$ with eigenvalue $E$. Then the matrix element
$\langle \psi_+|e^{i H \tau}|\psi_+\rangle$ can be trivially expressed in this eigenbasis as
\begin{equation}
\langle \psi_+|e^{i H \tau}|\psi_+\rangle= \sum_{E} |a_E|^2 e^{i E \tau}\, .
\label{matrix.1}
\end{equation}
Taking the absolute value square gives
\begin{equation}
|\langle \psi_+|e^{i H \tau}|\psi_+\rangle|^2= \sum_{E,E'} |a_E|^2\, |a_{E'}|^2\, e^{i (E-E')\,\tau}\, .
\label{matrix.2}
\end{equation}
Since the left hand side of Eq.~\eqref{matrix.2} is real, the imaginary part on the right hand side must vanish, leaving
\begin{equation}
|\langle \psi_+|e^{i H \tau}|\psi_+\rangle|^2=\sum_{E,E'}|a_E|^2\, |a_{E'}|^2\, \cos\left((E-E')\tau\right)\, .
\label{matrix.3}
\end{equation}
We now substitute this exact result [Eq.~\eqref{matrix.3}] in the integral in Eq.~(\ref{def_gr.1})
and use the identity
\begin{equation}
\int_0^{\infty}  d \tau\, e^{-r \tau}\, \cos\left( (E-E')\tau\right)= \frac{r}{r^2+(E-E')^2}\, .
\label{iden.1}
\end{equation}
This gives us an exact result for $\tilde{g}(r)$, valid for any $r$,
\begin{equation}
r\, \tilde{g}(r)=  \sum_{E,E'}|a_E|^2 |a_{E'}|^2\, \frac{r^2}{r^2 + (E-E')^2}\, .
\label{exact.1}
\end{equation}
We now take the limit $r\to 0$. In this limit, the ratio
\begin{equation}
{\displaystyle  \lim_{r\to 0}} \, \frac{r^2}{r^2+(E-E')^2}= \delta_{E,E'} \, ,
\label{limit.1}
\end{equation}
where $\delta_{E,E'}$ is the Kronecker delta function. Then we get from Eq.~(\ref{exact.1})
\begin{equation}
{\displaystyle \lim_{r\to 0}}\,  \left[r\, g(r)\right]= \sum_E |a_E|^4 \, .
\label{limit.2}
\end{equation}
Substituting this result in Eq.~(\ref{eq:ftd}) gives, as $r\to 0$,
\begin{equation}
\bar{t}_{r}\approx \frac{A_0}{r}\, , \quad {\rm where}\quad 
A_0= \frac{1}{\left[1- {\displaystyle \sum_E} |a_E|^4\right]}\, .
\label{smallr.1}
\end{equation}
This result is valid for any two-level Hamiltonian $H$. The prefactor $A_0$ 
can be computed exactly
for the Jaynes-Cummings model as we show in the next section (Sec.~\ref{sec:JC_mfd}).

Thus, the small and large $r$ behavior of the mean first detection time $\bar{t}_{r}$ in scheme
$1$ can be summarized as follows
\begin{eqnarray}
\bar{t}_{r}\approx \begin{cases}
& \frac{A_0}{r}\,  \quad\quad\quad\quad\quad\,\,\,\,\, {\rm as} \quad r\to 0\, , \\
&\\
&  \frac{r}{2\, |\langle \psi_+|H|\psi_{-}\rangle|^2 }\,\quad \quad {\rm as} \quad r\to \infty \, , 
\end{cases}
\label{meantr_asymp.1}
\end{eqnarray}
with $A_0$ given in Eq.~(\ref{smallr.1}). Thus, $\bar{t}_{r}$ diverges in the two opposite limits
$r\to 0$ and $r\to \infty$, provided the amplitude $A_0$ and the matrix element 
$\langle \psi_+|H|\psi_{-}\rangle$ are both nonzero.
In this case, $\bar{t}_{r}$ is clearly a non-monotonic function of $r$ and typically
has a unique minimum at some finite optimal value $r^*$.
This is somewhat reminiscent of the behavior of the mean first passage time to the origin of a 
Brownian motion on a line, starting at the initial position $x_0$
and resetting stochastically to $x_0$ with a constant rate $r$~\cite{EM11,EM12}.
However, the latter is a classical system and the physical mechanisms responsible for the emergence
of an optimal rate $r^*$ are quite different in quantum and classical systems. 
Note that in general the amplitude $A_0>0$ in Eq.~(\ref{meantr_asymp.1}) indicating
that the divergence $A_0/r$ is quite generic as $r\to 0$.
Hence the condition for the existence
of an optimal $r^*$ in scheme $1$ translates into the condition that the matrix element
$\langle \psi_+|H|\psi_{-}\rangle$ is finite. 
Since this is an off-diagonal element in a two-level system, it is nonzero provided the dynamics induced by
the Hamiltonian connects the two states. We will see in the next section (Sec.~\ref{sec:JC_mfd})
that in the Jaynes-Cummings model, this dynamics is provided by
the nonzero light-matter coupling $g$ that induces the flipping of the spin.\\

\subsection{Second scheme}
\label{eq:gen_second_scheme}

In the second scheme, we have $|\psi_{\rm int}\rangle= |\psi_+\rangle$
and hence $|\psi_c\rangle= |\psi_{-}\rangle$. Using this in
Eqs.~(\ref{eq:fdef}) and (\ref{eq:gdef}) gives
\begin{eqnarray}
\label{f_S2}
f (\tau) &\equiv&   \big| \langle   \psi_+  |  e^{i H \tau}  | \psi_{-} \rangle \big|^2\, , \\
g (\tau) &\equiv&    \big|   \langle \psi_{-}  \mid e^{i H \tau} | \psi_{-}  \rangle   \big|^2\, .
\label{g_S2}
\end{eqnarray}
Evidently, $f(\tau)\neq g(\tau)$ and hence
$\tilde{f}(s) \neq \tilde{g}(s)$. Consequently, the expression for $\tilde{S}_r(s)$
in Eq.~(\ref{eq:s_tilde_fg_laplace}) does not simplify further as in the first scheme
where $\tilde{f}(s)= \tilde{g}(s)$. However, it turns out that for the
second scheme, there exists another identity connecting $\tilde{f}(s)$ and $\tilde{g}(s)$ which reads
\begin{equation}
\label{eq:iden}
\tilde{f}(s) + \tilde{g}(s) = \frac{1}{s}\, .
\end{equation}
To prove this identity, we
add Eq.~(\ref{f_S2}) and Eq.~(\ref{g_S2})
and use the definition $|\psi_+\rangle\langle \psi_+|+ |\psi_{-}\rangle\langle \psi_{-}|=I$.
This gives
\begin{equation}
f(\tau)+g(\tau)=1\, . 
\label{eq:iden.1}
\end{equation}
Taking Laplace transform gives Eq. (\ref{eq:iden}).
Substituting Eq. (\ref{eq:iden}) in Eq.~\eqref{eq:s_tilde_fg_laplace} we get
\begin{equation}
\tilde{S}_r(s)= \frac{1}{(r+s)} \left[ 2- \frac{s}{(r+s) (1-r \tilde{g}(r+s))}\right]\, .
\label{eq_tilde_s_p2}
\end{equation}
Consequently,
\begin{equation}
\label{eq:ftilde_P2_gen}
\tilde{F}_r(s)= 1- s\,\tilde{S}_r(s) 
= 1 -\frac{2s}{(r+s)} - \frac{s^2}{(r+s)^2 (1-r \tilde{g}(r+s))}\, . 
\end{equation}

\vspace{0.4cm}

\noindent {\em Asymptotic behavior of $F_r(t)$ for small and large $t$.}
As in scheme $1$, let us now derive the small $t$ behavior of $F_r(t)$ by analysing
the large $s$ behavior of Eq.~(\ref{eq:ftilde_P2_gen}). For scheme $2$,
$|\psi_c\rangle= |\psi_{-}\rangle$, and consequently, from Eqs.~(\ref{eq:gdef}) and (\ref{eq:fg_laplace}),
we get 
\begin{equation}
\tilde{g}(s)= \int_0^{\infty} d\tau\, e^{-s\, \tau}\, \Big|\langle
\psi_- | e^{i H \tau}|\psi_-\rangle
\Big|^2\, .
\label{def_gr.P2}
\end{equation}
We then follow exactly the same analysis as in scheme $1$ to derive the large $s$ behavior
of $\tilde{g}(s)$, just by replacing $|\psi_+\rangle$ in Eq. (\ref{def_gr.1}) by $|\psi_{-}\rangle$.
The rest of the analysis goes through and we get exactly the same result as in 
Eq.~(\ref{eq:gen_g_large_s}),
where $\sigma^2$ happens to be the same as in Eq.~(\ref{sigma2.1}) since this is invariant
under the exchange $|\psi_+\rangle \to |\psi_{-}\rangle$.
We then substitute Eq.~\eqref{eq:gen_g_large_s} in Eq.~\eqref{eq:ftilde_P2_gen} and expand for large $s$.
This gives, to leading order, $\tilde{F}_r(s) \approx r/s$ and hence
\begin{equation}
F_r(t) \to r\, , \quad {\rm as} \quad t\to 0\, .
\label{smallr2.1}
\end{equation}
Thus, unlike in scheme $1$ where $F_r(t)$ vanishes as $t^2$ as $t\to 0$ in Eq.~(\ref{Frt_small.1}),
in this second scheme $F_r(t)$ approaches a constant $r$ as $t\to 0$. This drastic difference
in the small $t$ behavior in the two schemes can be easily understood as follows. 
As $t\to 0$, the system has hardly evolved
from its initial state $|\psi_+\rangle$. So, if a measurement takes place (with probability $r\, dt$ in
a small time $dt$), the state of interest will be definitely detected if it coincides with
the initial state $|\psi_+\rangle$ and this is exactly the case in scheme $2$. Hence, the probability
that a detection occurs in a time interval $[0,dt]$ is simply the probability that a measurement takes place
in $[0,dt]$ and this occurs with probability $r\,dt$. Hence $F_r(t)\to r$ as $t\to 0$ in 
scheme $2$.
In contrast, in scheme $1$, the state of interest $|\psi_-\rangle$ differs from the 
initial state $|\psi_+\rangle$
and the probability of its detection $[0,dt]$ vanishes as $t\to 0$.

The large $t$ behavior of $F_r(t)$, as in scheme 1, will be typically exponential and the 
inverse time scale of the exponential decay is fixed by the pole of $\tilde{F}_r(s)$ in Eq.~(\ref{eq:ftilde_P2_gen}) in the complex $s$ plane.  It clearly depends on the details of 
$\tilde{g}(s)$ and an explicit example will again be worked out in the next section for the 
Jaynes-Cummings model.

\vspace{0.4cm}

\noindent {\em Mean first detection time.}
The mean first detection time in the second scheme, setting $s=0$ in Eq.~(\ref{eq_tilde_s_p2}), is given by
\begin{equation}
\bar{t}_r =  \tilde{S}_r (0)  =  \frac{2}{r}\, .
\label{eq:ftdneq}
\end{equation}
This is again strikingly different from the first scheme.
In particular, we note that $\bar{t}_{r}$ in Eq.~(\ref{eq:ftdneq}) 
decays monotonically with increasing $r$. This is at variance
with scheme $1$, where $\bar{t}_{r}$ diverges in both limits $r\to 0$
and $r\to \infty$ quite generically, implying the existence of an optimal
$r^*$ at an intermediate value. In scheme $2$, there is no optimal $r^*$.
The fact that $\bar{t}_{r}$ vanishes as $r\to \infty$ limit in scheme $2$
can be understood very easily, once again using the implication of the Zeno limit. 
When $r\to \infty$, detection
attempts take place very frequently and
the state stays put in its initial state $|\psi_+\rangle$ at all times.
So, in scheme $2$ where the state of interest
$|\psi_{\rm int}\rangle$ coincides with the initial state $|\psi_+\rangle$, it will
be detected immediately at $t=0$ (with probability $1$) when $r$ is large. Consequently,
the mean detection time must vanish as $r\to \infty$ in scheme $2$, in contrast
to scheme $1$ where it diverges as $r\to \infty$. 
This result in Eq.~\eqref{eq:ftdneq} already appeared in Refs.~\cite{KEZ21, ZBK21} 
using a different method, and the `topological' significance of the factor `2' in Eq. (\ref{eq:ftdneq})
was pointed out (see also section~\eqref{sec:general} for a discussion of this
prefactor).

The results presented in this section hold for a general Hamiltonian that is characterized by two levels. 
Next, in Sec.~\ref{sec:JC_mfd}, we will derive explicitly
the full first detection probability $F_r(t)$ and the mean detection time $\bar{t}_r$
in a specific two-level system. This is the Jaynes-Cummings
Hamiltonian which plays a fundamental role in
hybrid quantum systems.

\section{Explicit first detection probability in the Jaynes-Cummings model}
\label{sec:JC_mfd}

In this section, we derive explicit results for $F_r(t)$ and its moments
for both schemes in the Jaynes-Cummings (JC) model~\cite{GSA13,SZ97,C99,HR06}. We start by recalling
briefly this model.

In the JC model a single spin $1/2$ or a qubit (matter) is coupled to a single 
cavity mode (light) via the Hamiltonian
\begin{equation}
H = \omega_q\, S^z + \omega_c\, a^{\dagger}\, a + g\, [S^{+} a + a^{\dagger} S^-]\, ,
\label{eq:main_JCHam}
\end{equation}
where $\omega_q$ is the qubit frequency, $\omega_c$ is the cavity frequency and $g$ is the light-matter 
coupling. Here $a / a^{\dagger}$ are annihilation/creation operators of bosons and $S^{\pm}, S^z$ are the
Pauli spin operators. The model [Eq.~\eqref{eq:main_JCHam}] is defined on a direct product of two spaces, i.e., a 
two level Hilbert space (spanned by $\mid \uparrow \rangle, \mid\downarrow \rangle$) and a single mode 
cavity Hilbert space (spanned by $\big| n \rangle$ where $n = 0,1,2,3 \dots$). One can define an excitation 
operator $\hat{N} = a^{\dagger} a + S^+ S^-$ which commutes with the Hamiltonian $H$. 
For a given excitation sector labelled by $n$, the Hilbert space is spanned by 
only two states $\mid \downarrow, n \rangle$ and 
$\mid \uparrow, n-1 \rangle$. 
Since $\hat N|\mid \downarrow, n \rangle=n\, \mid \downarrow, n \rangle$
and $\hat N|\mid \uparrow, n-1 \rangle=n\, \mid \uparrow, n-1 \rangle$,
both these states have the 
same total excitation $n$ and the Hamiltonian 
does not couple spaces spanned by different excitation sectors.
Therefore, given a fixed excitation sector labelled by $n$,
we have effectively a two level system consisting of two states
$\mid \downarrow, n \rangle$ and $\mid \uparrow, n-1 \rangle$. This makes 
the JC Hamiltonian a putative platform for investigating the probability distribution
of the first detection time in a two level system.

The general notation $| \psi_+\rangle$ and $| \psi_-\rangle$ given in Table.~\ref{tab_table1} and discussed 
in Sec.~\ref{sec:GE} now stand for $\mid \uparrow, n-1 \rangle$ and $\mid \downarrow, n \rangle$ 
respectively. For the sake of completeness, a detailed description of the JC Hamiltonian and its exact 
solvability structure is discussed in ~\ref{app_JC}. We will now present results for 
both schemes represented in Table~\ref{tab_table1}. For the simplicity of presentation, we restrict ourselves
to the resonant limit where $\omega_q=\omega_c$. The case $\omega_q\ne \omega_c$ can also be
treated, though the results are cumbersome and not illuminating. Hence we restrict
ourselves to $\omega_q=\omega_c$ where the formulae become simpler.

\subsection{First scheme for the JC Hamiltonian: }
\label{subsec:p1}
We start the system from the initial state 
\begin{equation}
\label{eq:JC_psi0}
 | \psi(0) \rangle = | \psi_+\rangle =   \mid \uparrow, n-1 \rangle \, ,
\end{equation}
and recall that in this scheme, 
the state of interest (detection) is different from the initial state, i.e., 
\begin{equation}
\label{eq:JC_init1}
|\psi_{\rm int} \rangle  =  | \psi_-\rangle =   \mid \downarrow, n \rangle \, . 
\end{equation}
Consequently, $|\psi_c\rangle=  | \psi_+\rangle = \mid \uparrow, n-1 \rangle$. 

To compute the Laplace transform of $F_r(t)$ given in Eq.~(\ref{eq:ftilde_P1_gen}),
we need to first compute $\tilde{g}(r+s)$. This can be done explicitly for
the JC model, as illustrated below. 
In order to compute $\tilde{g}(s)$ in Eq.~\eqref{def_gr.1}, 
we need to evaluate $e^{i H \tau} | \psi_+ \rangle = e^{i H \tau}  \mid \uparrow, n-1 \rangle$. 
In ~\ref{Initial_JC} we evaluate this and obtain
\begin{equation}
\label{eq:JCevol}
 e^{i H \tau} | \psi_+ \rangle = e^{i H \tau} \mid \uparrow, n-1 \rangle 
 = e^{i \omega_c\big(n-1/2 \big)\, \tau} \Big[-i \sin(g\sqrt{n}\, \tau ) \mid \downarrow, n 
\rangle    
 + \cos(g\sqrt{n} \tau) \mid \uparrow, n-1 \rangle   \Big] \, . 
\end{equation}
It then follows that
\begin{equation}
\label{eq:JCevol_prod}
\Big| \langle   n-1,\uparrow \mid  \, e^{i H \tau}  \, \mid \uparrow , n-1 \rangle   \Big|^2 = 
\cos^2 (g\,\sqrt{n}\, \tau) \, .
\end{equation}
Substituting this result in Eq.~\eqref{def_gr.1} gives
\begin{equation}
\label{eq:grs0_JC}
\tilde{g}(s) =  \int_0^{\infty} d\tau  e^{-s\tau}   \big|   \langle n-1,\uparrow \mid 
e^{i H \tau} \mid \uparrow , n-1 \rangle   \big|^2 
=  \frac{1}{s}\, \frac{2 g^2 n+s^2}{4 g^2 n +s^2}
\end{equation}
Subsequently, from Eqs.~\eqref{eq:s_tilde_fg_laplace_simp} and \eqref{eq:ftilde_P1_gen}, we get 
\begin{equation}
\tilde{S}_r (s)
=  \frac{(r+s)^2+4g^2n}{s(r+s)^2+2g^2 n r+4 g^2 n s} \, ,
\label{eq:s_tilde_fg_laplace_simp_eg1}
\end{equation}
and
\begin{equation}
\tilde{F}_r(s) 
=   \frac{2g^2 n r}{s(r+s)^2+2g^2 n r+4 g^2 n s} \, .
\label{eq:tildeFexp}
\end{equation}
Inverting the Laplace transform in Eq.~(\ref{eq:tildeFexp}) we get 
\begin{equation}
F_r(t) = \int_{\Gamma} \frac{ds}{2\pi i }\, e^{s\,t}\, \bigg[\frac{2g^2 n r}{s(r+s)^2+2g^2 n r+4 g^2 n s} 
\bigg]\, ,
\label{eq_inv_lt}
\end{equation}
where $\Gamma$ denotes a Bromwich contour in the complex $s$ plane. It turns out to
be convenient to rescale $s= r\, \lambda$ and define the dimensionless parameter
\begin{equation}
\label{eq:cov}
\mu = \frac{2\,g^2\, n}{r^2} \, . 
\end{equation}
Upon this rescaling in Eq.~\eqref{eq_inv_lt}, one finds that $F_r(t)$ can be expressed 
in the scaling form
\begin{equation}
F_r(t) = r\,G^{(1)}_\mu(r t) \, ,
\label{eq_inv_lt2}
\end{equation}
with
\begin{equation}
G^{(1)}_\mu(z)= 
\int_{\Gamma} \frac{d\lambda}{2\pi i }\, e^{\lambda\, z}\, 
\bigg[\frac{\mu}{\lambda^3+2\lambda^2+\lambda(1+2\mu)+\mu} \bigg]\, ,
\label{eq_inv_lt1}
\end{equation}
and the superscript $(1)$ stands for scheme $1$.
The integrand has three poles $(\lambda_1,\, \lambda_2,\, \lambda_3)$ in the complex $\lambda$ plane,
all with negative real parts. These are the three roots of the cubic equation
\begin{equation}
\label{eq:cubic}
\lambda^3+2\, \lambda^2+\lambda\, (1+2\,\mu)+\mu = 0 \, ,
\end{equation}
and can be explicitly expressed as
\begin{equation}
\label{eq_cub_sol}
\lambda_k = -\frac{1}{3} \bigg(2+\zeta^k C + \frac{B_0}{\zeta^k C} \bigg)\,\quad \text{for }\quad k=1,\,2,\,3
\end{equation}
where 
\begin{equation}
\zeta = \frac{-1+i\sqrt{3}}{2}, \quad B_0=1-6\mu ,\quad 
C = \sqrt[3]{\frac{B_1 + \sqrt{B_1^2- 4 B_0^3}}{2}},\quad B_1 = -2 - 9\mu.
\label{C_def}
\end{equation}
The real parts of all three roots in Eq.~\eqref{eq_cub_sol} 
are negative. 
Out of the three roots, $\lambda_1$ is real negative and the other two are complex conjugate pairs
($\lambda_3 = \lambda_2^*$) with their common real part smaller than $\lambda_1$.
The Bromwich integral in Eq.~(\ref{eq_inv_lt1}) can then be performed explicitly and 
is given by the sum of the residues at these
three poles. One gets 
\begin{equation}
G^{(1)}_\mu(z) = \mu \bigg[  \frac{e^{\lambda_1 z}}{(\lambda_1-\lambda_2)(\lambda_1-\lambda_3)} + 
\frac{e^{\lambda_2 z}}{(\lambda_2-\lambda_1)(\lambda_2-\lambda_3)}  
+    \frac{e^{\lambda_3 z}}{(\lambda_3-\lambda_1)(\lambda_3-\lambda_2)} \bigg] \, .
\label{eq_inv_lt3}
\end{equation}
Using $\lambda_3=\lambda_2^*$, one can further simplify Eq.~\eqref{eq_inv_lt3} and
express the scaling function $G^{(1)}_\mu(z)$ in terms of only real variables. 
Defining
$\lambda_2 = \lambda_R+i\, \lambda_I$ and $\lambda_3 = \lambda_R - i\, \lambda_I$, where
$\lambda_R<\lambda_1<0$ and $\lambda_I>0$ are real, we get from Eq.~(\ref{eq_inv_lt3})
\begin{equation}
G^{(1)}_\mu(z) =\mu \bigg[  \frac{e^{\lambda_1 z}}{(\lambda_1-\lambda_R)^2 +\lambda_I^2} 
-    \frac{e^{\lambda_R z} \big( (\lambda_1 - \lambda_R) \sin(\lambda_I z) + 
\lambda_I \cos(\lambda_I z)\big)}{\lambda_I\big((\lambda_1-\lambda_R)^2\ + \lambda_I^2 \big)} 
\bigg] \, .
\label{eq_inv_lt3_trig}
\end{equation}
This result clearly illustrates that $F_r(t)= r\, G^{(1)}_\mu(rt)$ has two parts: the first part
decays purely exponentially $e^{-|\lambda_1|\, r\, t}$ at all times, while the second part has
oscillatory terms of period $2\pi/{r \lambda_I}$, but the amplitude of the oscillations dies out 
exponentially as $e^{-|\lambda_R|\, r\,t}$.
From this explicit formula, we see that there are three time scales associated with the temporal
evolution of $F_r(t)$, given respectively by
$t_1=1/|r\, \lambda_1|$, $t_2=1/|r\, \lambda_R|$ and $t_3= 2\pi/{r\,\lambda_I}$.
Of the three time scales, $t_3$ is the shortest.
Since $|\lambda_R|> |\lambda_1|$, we have $t_2<t_1$ and hence
the late time exponential decay of $F_r(t)$ is solely governed
by the largest time scale $t_1$. We will remark more on this important time scale later.

\begin{figure}
\centering
		\includegraphics[width=0.7\linewidth]{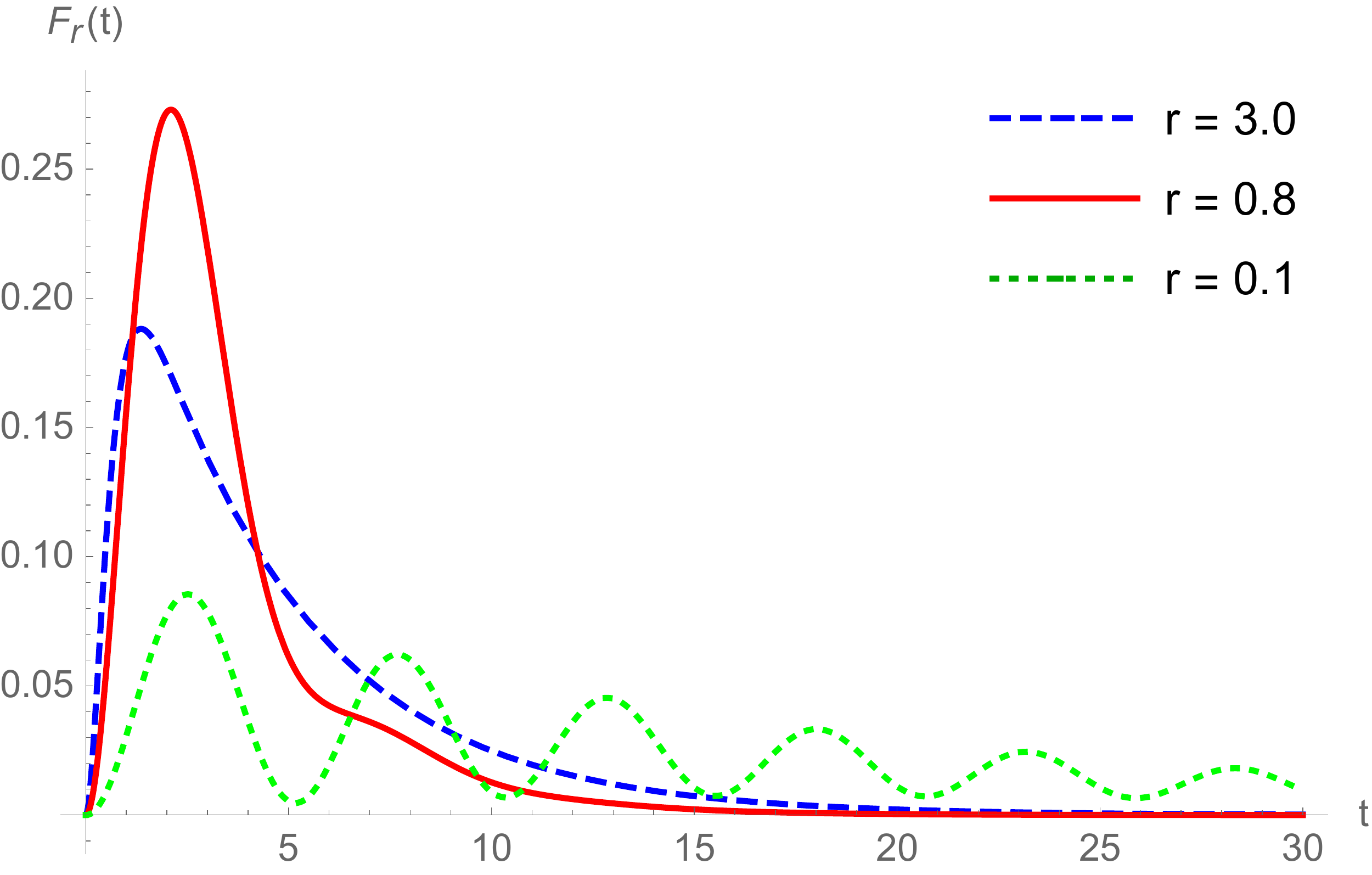}
		\caption{For the first scheme (Table.~\ref{tab_table1}), the PDF $F_r(t)$ 
in Eq.~(\ref{eq_inv_lt2}), with $G^{(1)}_\mu(z)$ in Eq.~(\ref{eq_inv_lt3_trig}),
is plotted as a 
function of $t$ for various values of $r$. Blue (dashed), red (solid), and green (dotted) 
correspond to $r = 3.0, 0.8$ and $0.1$ respectively. 
We take light-matter coupling $g = 0.1$ and excitation sector $n = 37$. }
	\label{fig:p1frt}
\end{figure}

From the exact expression of $F_r(t)= r\, G^{(1)}_\mu(rt)$, one can easily work out its 
asymptotic behaviors for small and large $t$. We get 
\begin{eqnarray}
F_r(t) \approx \begin{cases}
& r\, g^2\, n\, t^2\, \quad\quad\quad\quad\quad\quad\quad\, {\rm as}\quad t\to 0 \\
&\\
& \frac{r\,\mu}{\left[(\lambda_1-\lambda_R)^2 + \lambda_I^2\right]}\, e^{-|\lambda_1|\, r\, t}\ \quad 
{\rm as} \quad  t\to \infty\, ,
\end{cases}
\label{Fr1_asymp}
\end{eqnarray}
where $\lambda_1$, $\lambda_R$ and $\lambda_I$ can be read off Eq.~(\ref{eq_cub_sol}). We note
that the small $t$ behavior in Eq.~(\ref{Fr1_asymp}) is in complete agreement with the small $t$
behavior derived in Eq.~(\ref{Frt_small.1}) for
a general two level system. This can be verified by computing $\sigma^2$ in Eq.~(\ref{sigma2.1}) for
the JC model. We get $\sigma^2= g^2 n$ (see Eq.~(\ref{eq:hn}) where the off-diagonal
element is precisely $\sigma$).
In Fig.~\ref{fig:p1frt}, we plot $F_r(t)$ vs. $t$ for different values of $r$. One clearly
sees oscillations for small values of $r$, but they get suppressed as $r$ increases.
Similarly, for fixed $r$, the oscillations get suppressed for large $t$.

The mean first detection time $\bar{t}_r $, setting $s=0$ in Eq.~\eqref{eq:s_tilde_fg_laplace_simp_eg1}, reads
\begin{equation}
\bar{t}_r = \tilde{S}_r (0) =  \frac{2}{r} +\frac{r}{2g^2 n}\, .
\label{eq:ftd_JC1}
\end{equation}
As a function of $r$, the mean first detection time diverges in the two extreme limits $r\to 0$ and $r\to \infty$.
One can easily check that the small $r$ behavior $\approx 2/r$ in Eq.~(\ref{eq:ftd_JC1})
matches perfectly with the general small $r$ asymptotics in Eq.~(\ref{smallr.1}) valid
for arbitrary two-level system. Indeed, for the JC model it is easy to show that
$a_E=1/\sqrt{2}$ for
both the energy eigenstates. Hence, in this case, ${\displaystyle \sum_E} |a_E|^4= 1/2$. Consequently,
from Eq.~(\ref{smallr.1}), we get $A_0=2$ which is in perfect agreement with the exact result
in Eq.~(\ref{eq:ftd_JC1}) as $r\to 0$. Similarly, the large $r$ behavior $\approx \frac{r}{2g^2 n}$
in Eq.~(\ref{eq:ftd_JC1}) is also consistent with the general large $r$ asymptotic in
Eq.~(\ref{tr_final.1}). To check this, we note that for scheme $1$, 
$| \psi_+\rangle = \mid \uparrow, n-1 \rangle$ and 
$|\psi_{-}\rangle= \mid \downarrow, n \rangle$. 
The matrix element $\langle \psi_+|H|\psi_{-}\rangle$ appearing in Eq.~(\ref{tr_final.1})
can then be computed for the JC model and is simply given by $g\, \sqrt{n}$ (see Eq.~(\ref{eq:hn}) in
~\ref{app_JC}).
Consequently, Eq.~(\ref{tr_final.1}) is fully consistent with the exact large $r$ behavior
in Eq.~(\ref{eq:ftd_JC1}).

\begin{figure}
\centering
		\includegraphics[width=0.7\linewidth]{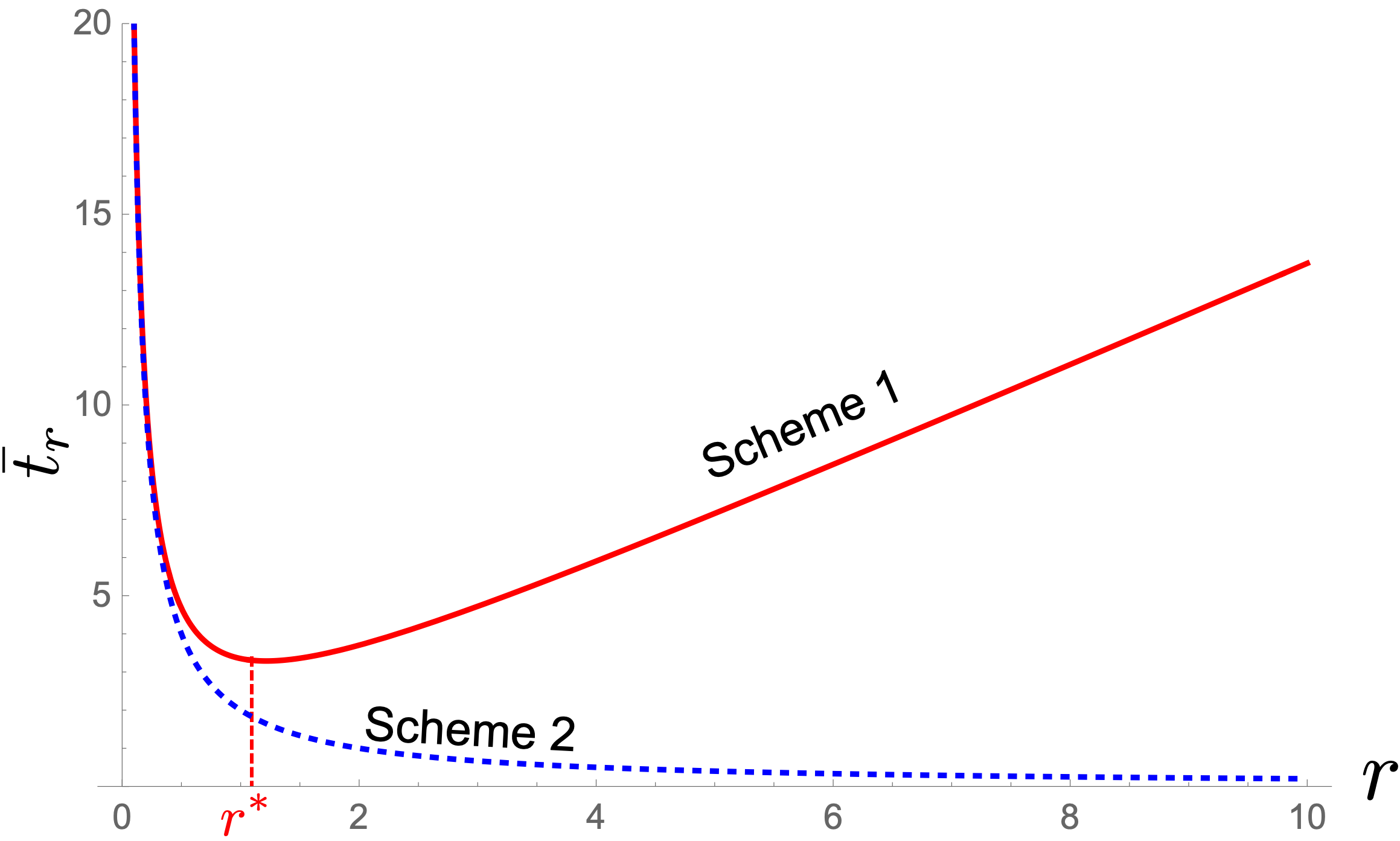}
		\caption{The mean first detection time $\bar{t}_r$ plotted as a function of the measurement rate $r$, for
schemes $1$ and $2$ 
[Eq.~\eqref{eq:ftd_JC1} for scheme $1$ and Eq.~\eqref{eq:ftd_JC1_P2} for scheme $2$]. 
In scheme $1$, $\bar{t}_r$ behaves non-monotonically with increasing $r$ with a unique minimum
at $r=r^*$, while in scheme $2$ it decreases monotonically with increasing $r$.
The location of $r^*$ for scheme $1$ 
[Eq.~\eqref{eq:r_opt}] is also shown. For scheme $1$, we take $g = 0.1$ and $n = 37$. For scheme $2$,
$\bar{t}_r$ is independent of $g$ and $n$.}
	\label{fig:mean}
\end{figure}

This non-monotonic dependence of $\bar{t}_{r}$ on $r$ is structurally similar to the non-monotonic behavior
of the mean first passage time of a classical Brownian motion in the presence
of a stochastic resetting with rate $r$~\cite{EM11,EM12}.
The mean $\bar{t}_{r}$ in Eq.~(\ref{eq:ftd_JC1}), as a function of $r$,
has a single unique minimum as $r=r^*$ for which the mean detection time is minimal/optimal. 
The value of $r^*$ is given by
\begin{equation}
\label{eq:r_opt}
r^{*} = 2\,g\, \sqrt{n}\, ,
\end{equation}
where we recall that $g$ is the light-matter 
coupling and $n$ is the label of the fixed excitation sector. 
At this optimal resetting rate, the value of mean first detection time is 
\begin{equation}
\label{r_opt.1}
\bar{t}_{r=r^*}  =  \frac{2}{g\,\sqrt{n}} \, . 
\end{equation}
In Fig.~\ref{fig:mean}, a plot of $\bar{t}_r$ vs $r$ [Eq.~\eqref{eq:ftd_JC1}] is shown (red solid line), where the 
location of the optimal resetting rate [Eq.~\eqref{eq:r_opt}] is also marked.

Similarly, the second moment is given by
\begin{equation}
\mu_2 =  \int_0^{\infty} F_r(t)\, t^2\, dt = -2 \tilde{S}^{\prime}_r (s=0) 
=  2\bigg[\frac{1}{a^2} +\frac{4}{r^2} + \frac{r^2}{4 a^4} \bigg] \, ,
\end{equation}
where $a = g\sqrt{n}$. The variance is given by 
\begin{equation}
\label{eq:fluc}
\sigma_{\rm fd}^2 = \mu_2 - \mu_1^2 
= \frac{4}{r^2} + \frac{r^2}{4a^4}\, , 
\end{equation}
where the subscript `${\rm {fd}}$' stands for `first detection'.
Interestingly, it turns out that the variance  in Eq.~(\ref{eq:fluc}) also has a 
minimum exactly at the same value of $r^*$ [Eq.~\eqref{eq:r_opt}], as for the mean [Eq.~\eqref{eq:ftd_JC1}].
Thus, not only the mean of the first detection time becomes minimal at $r=r^*$, its fluctuations
around the mean are also minimal when $r=r^*$.

\begin{figure}
\centering
		\includegraphics[width=0.7\linewidth]{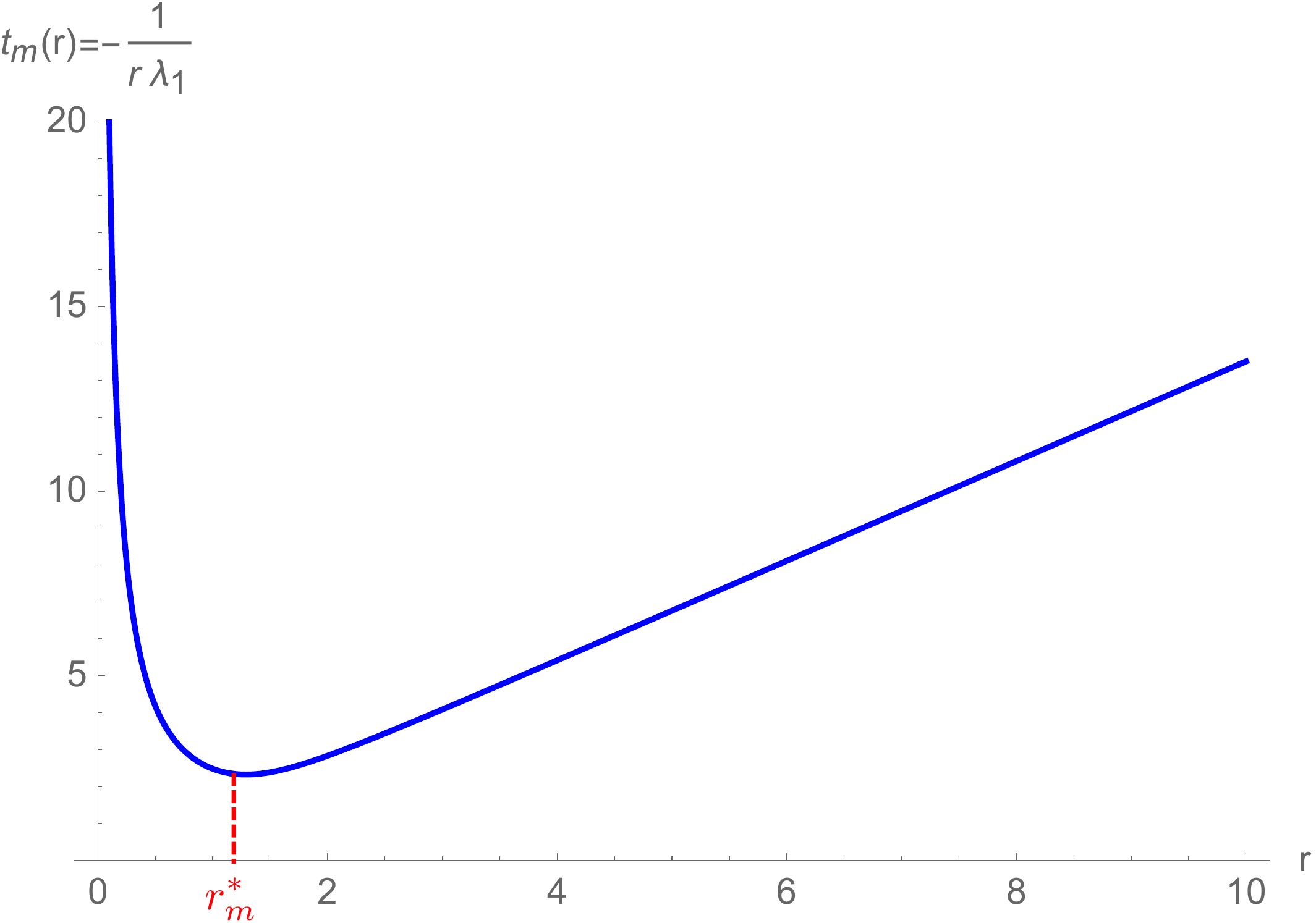}
		\caption{Plot of the maximum waiting time $t_m(r)$ vs. 
$r$ [Eq.~\eqref{eq:tm}] and also the location of 
the optimal resetting rate $r_m^*$. We take $g = 0.1$ and $n = 37$. }
\label{fig:tm}
\end{figure}

Let us conclude this subsection with a final remark. The first moment 
$\bar{t}_r $ is, by definition, the average amount of time taken for a signal to be detected for the 
first time. However, the average time may not always be representative of different time scales
present in the dynamics. For example, from the exact solution of $F_r(t)$ in Eq.~(\ref{eq_inv_lt3_trig}),
we have seen the existence of three different time scales. In particular, the longest time scale
$t_1=1/|r\, \lambda_1|$ governs the decay of $F_r(t)$ at late times [see the second line of 
Eq. (\ref{Fr1_asymp})]. The physical meaning of this time scale is as follows: it is
clear from Eq.~(\ref{Fr1_asymp}) that for $t>>t_1$, the first detection probability $F_r(t)$
is essentially zero. So, $t_1$ is effectively the `maximal' time one needs to wait to detect a signal with 
a nonzero probability. When $t$ exceeds $t_1$, the detection probability vanishes almost surely.
It is then appropriate to denote this time scale by $t_m(r)$, with the subscript $m$
standing for `maximal', i.e.,
\begin{equation}
\label{eq:tm}
t_m(r) = t_1= -\frac{1}{r\lambda_1} \, .
\end{equation}
From the exact expression of $\lambda_1$ in Eqs.~(\ref{eq_cub_sol}) and (\ref{C_def}),
one can show that $t_m(r)$ is again a non-monotonic function of $r$, diverging in the two
opposite limits, as
\begin{eqnarray}
t_m(r) \approx \begin{cases}
& \frac{2}{r} \quad\quad\,\,\, {\rm as} \quad r\to 0  \\
& \frac{r}{2\,g^2\,n} \quad {\rm as} \quad r\to \infty \, .
\end{cases}
\label{tmr_asymp}
\end{eqnarray}
Thus, interestingly, the asymptotic behaviors of $t_m(r)$ are exactly similar to that of the
mean $\bar{t}_{r}$ in Eq.~(\ref{eq:ftd_JC1}). However, they do differ
from each other for intermediate values of $r$. In Fig.~\ref{fig:tm}, we plot
$t_m(r)$ vs. $r$ which also exhibits a unique minimum at some $r=r_m^*$. However, this
optimal value differs from the optimal value $r^*$ that minimizes $\bar{t}_{r}$. Nevertheless,
the fact remains that there is always a certain optimal detection rate $r$ that minimizes the
time of detection in scheme $1$, no matter which time scale one uses. In the next
subsection (Sec.~\ref{subsec:p2}), we will show that even for scheme $2$ where the mean $\bar{t}_{r}$ decreases
monotonically with increasing $r$, the time scale $t_m(r)$ is identical to that of scheme $1$.
Hence, choosing $r=r_m^*$ minimizes the maximal detection time scale $t_m(r)$ in
both schemes.

\begin{figure}
\centering
		\includegraphics[width=0.7\linewidth]{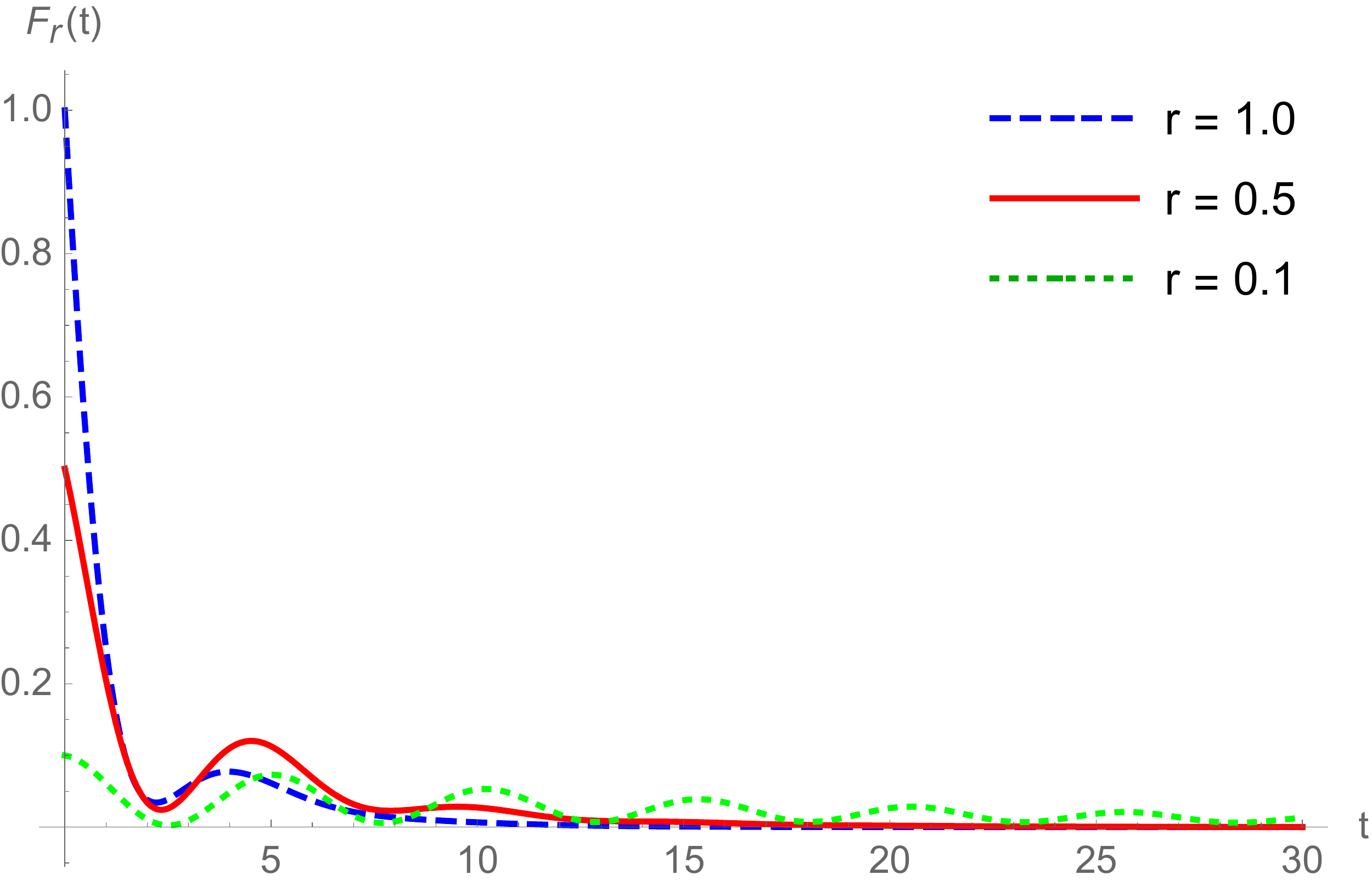}
		\caption{For the second scheme (Table.~\ref{tab_table1}), 
the distribution of the first detection time $F_r(t)$ [Eq.~\eqref{eq_inv_lt2_P2}] is plotted as a 
function of time $t$ for 
various values of $r$. Blue (dashed), red (solid), and green (dotted) correspond to $r = 1.0, 0.5$ and $0.1$ respectively. We chose 
$g = 0.1$ and $n = 37$. }
	\label{fig:p2frt}
\end{figure}

\subsection{Second scheme for the JC Hamiltonian: }
\label{subsec:p2}

Here again we start the system in the same initial state as in scheme $1$ in
Eq.~(\ref{eq:JC_psi0}), but now the state of interest 
$|\psi_{\rm int} \rangle$ is given by
\begin{equation}
\label{eq:JC_init1_P2}
|\psi_{\rm int} \rangle  =  | \psi_+\rangle =    \mid \uparrow, n-1 \rangle\, . 
\end{equation}
Consequently, the complementary state in this case is  
$|\psi_c\rangle=  | \psi_-\rangle = \mid \downarrow, n \rangle$.  Again, to
compute the Laplace transform of $F_r(t)$ in Eq.~(\ref{eq:ftilde_P2_gen}),
we need to compute $\tilde{g}(s)$
in Eq.~(\ref{def_gr.P2}). The matrix element $\langle
\mid \downarrow, n | e^{i H \tau} \mid \downarrow, n\rangle$ can again be easily evaluated 
(see ~\ref{app_JC}) and one finds that
\begin{equation}
\tilde{g}(s) =  \int_0^{\infty} d\tau\,  e^{-s\tau}\,   \big|   
\langle \downarrow, n \mid e^{i H \tau} \mid \downarrow, n \rangle   \big|^2 
= \frac{1}{s}\, \frac{2 g^2 n+s^2}{4 g^2 n +s^2} \, .
\label{eq:fg_laplace_recall_neq1}
\end{equation}
Consequently, from Eqs.~\eqref{eq_tilde_s_p2} and \eqref{eq:ftilde_P2_gen}, we get 
\begin{equation}
\tilde{S}_r (s) =\frac{4 g^2 n+s (r+s)}{2 g^2 n (r+2 s)+s (r+s)^2} \, ,
\label{eq:srs_neq}
\end{equation}
and  
\begin{equation}
\tilde{F}_r(s) 
=  \frac{r \left(2 g^2 n+s (r+s)\right)}{2 g^2 n (r+2 s)+s (r+s)^2} .
\label{eq:tildeFexp_P2}
\end{equation}

\begin{figure}
\centering
		\includegraphics[width=0.7\linewidth]{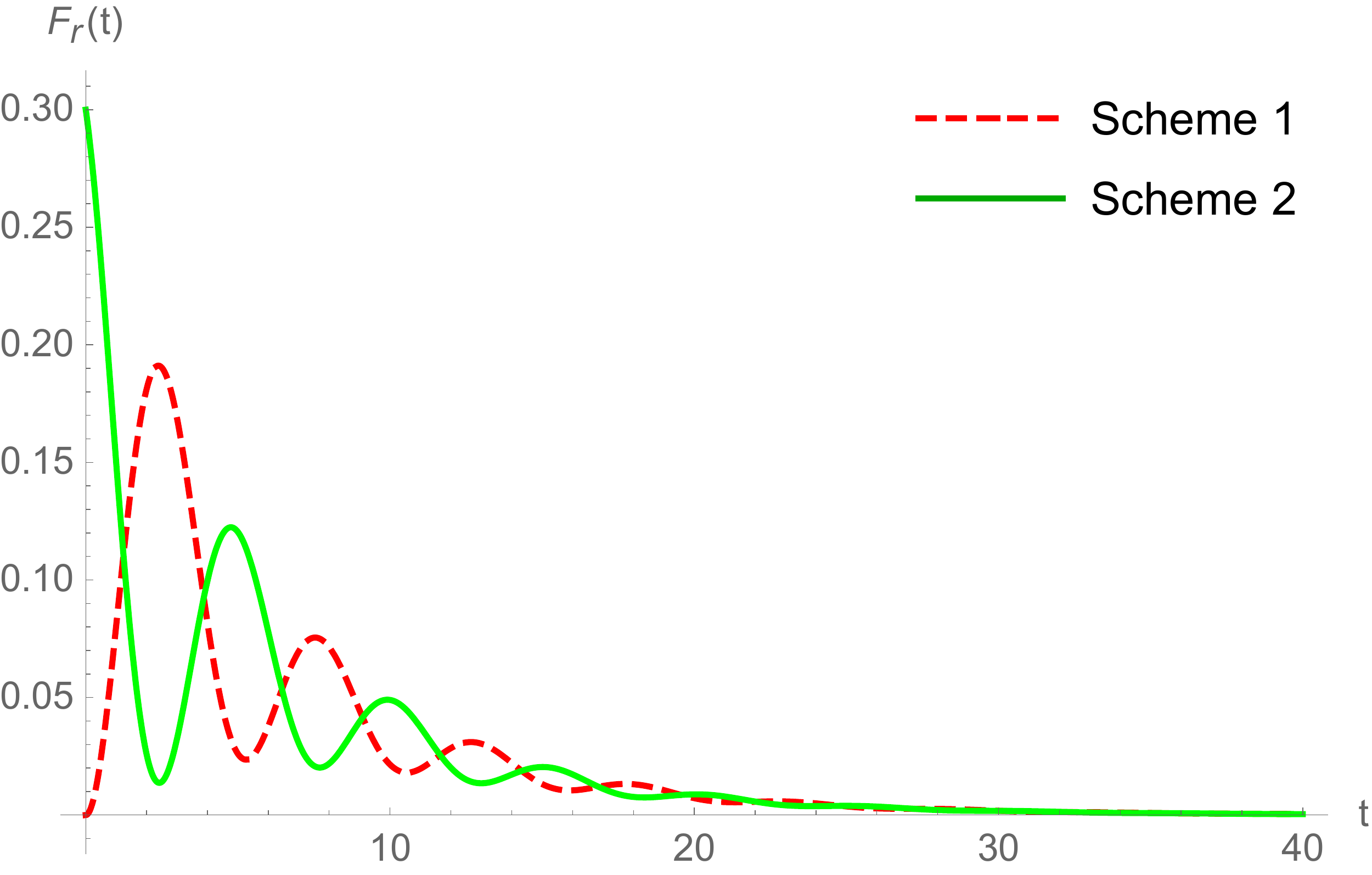}
		\caption{
The PDF of the first detection time $F_r(t)$ plotted as a function of $t$ for
scheme $1$ [Eq.~(\ref{eq_inv_lt2})] and scheme $2$ [Eq.~\eqref{eq_inv_lt2_P2}]. 
We chose  $r=0.3, g = 0.1$ and $n = 37$. }
	\label{fig:p2frt_com}
\end{figure}

We next follow the same steps as in Sec.~\ref{subsec:p1} including the same change of variables. We get,
\begin{equation}
F_r(t) = r\,G^{(2)}_\mu(r t),
\label{eq_inv_lt2_P2}
\end{equation}
where 
\begin{equation}
G^{(2)}_\mu(z) =  \int_{\Gamma} 
\frac{d\lambda}{2\pi i \,}e^{\lambda\, z}\, 
\bigg[\frac{\mu+\lambda(\lambda+1)}{\lambda^3+2\lambda^2+\lambda(1+2\mu)+\mu} \bigg] \, .
\label{eq_inv_lt1_P2}
\end{equation}
Note that the integrand in Eq.~(\ref{eq_inv_lt1_P2}) has the identical three poles $\lambda_1$,
$\lambda_2$ and $\lambda_3$ as in Sec.~\ref{subsec:p1}.
Evaluating the residues at these poles, one gets
\begin{equation}
G^{(2)}_\mu(z) =  \frac{\big(\mu+\lambda_1(\lambda_1+1)\big)e^{\lambda_1 z}}{(\lambda_1-
\lambda_2)(\lambda_1-\lambda_3)}+ \frac{\big(\mu+\lambda_2(\lambda_2+1)\big)
e^{\lambda_2 z}}{(\lambda_2-\lambda_1)(\lambda_2-\lambda_3)}  
+ \frac{\big(\mu+\lambda_3(\lambda_3+1)\big)e^{\lambda_3 z}}{(\lambda_3-\lambda_1)(\lambda_3-\lambda_2)}, 
\label{eq_inv_lt3_P2}
\end{equation}
where $\lambda_1$, $\lambda_2$ and $\lambda_3$ are the roots of the cubic 
equation [Eq.~\eqref{eq:cubic}]. 
As in Sec.~\ref{subsec:p1},  one can express the scaling function
in terms of the real variables $\lambda_1$, $\lambda_R$ and $\lambda_I$
which makes the oscillations manifest. But the expression is a bit long
and cumbersome, so we omit it here. From the exact result for $F_r(t)$ in scheme $2$, one can
easily extract its asymptotic behaviors
for small and large $t$, with $r$ fixed. One finds
\begin{eqnarray}
F_r(t) \approx \begin{cases}
& r\, \quad\quad\quad\quad\quad\,\, {\rm as}\quad t\to 0  \\
& A_{2}\, e^{-|\lambda_1|\, r\, t}\ \quad {\rm as} \quad  t\to \infty\, ,
\end{cases}
\label{Fr2_asymp}
\end{eqnarray}
where the amplitude $A_2$ can be computed in terms of the three roots $\lambda_k$'s.
We plot $F_r(t)= r\, G^{(2)}_\mu(rt)$ as a function of $t$ for different values of $r$
in Fig.~\ref{fig:p2frt}. Note that $F_r(t)$ 
in the second scheme is drastically different from that in the first scheme 
(Sec.~\ref{subsec:p1}), especially at small times $t$. 
From Eq.~\eqref{eq_inv_lt2_P2}, it is easy to see that 
$F_r(0) = r$ (i.e., a constant). This is also evident in the large-$s$ behaviour for 
Eq.~\eqref{eq:tildeFexp_P2} which 
gives $\tilde{F}_r(s) \approx r/s$ as $s\to \infty$. 
In Fig.~\eqref{fig:p2frt_com}, we compare $F_r(t)$ vs. $t$ in the two schemes.

Setting $s=0$ in Eq.~\eqref{eq:srs_neq}, the mean detection time for the second scheme becomes
\begin{equation}
\bar{t}_r = \tilde{S}_r (0) =  \frac{2}{r} 
\label{eq:ftd_JC1_P2}
\end{equation}
This result shows for the second scheme, there is no finite optimal resetting rate $r^*$ 
at which the mean first detection time is minimum. In other words, $r^*$ is trivially $\infty$.
This is quite different from the behavior of $\bar{t}_{r}$ in scheme $1$ which does have
a minimum at a finite $r^*$. 
In Fig.~\ref{fig:mean}, we show the mean first detection 
time [Eq.~\eqref{eq:ftd_JC1_P2}] for the second scheme (blue dotted line), where
it is compared to that of the scheme $1$ (red solid line). 

While the mean detection time is very different in the two schemes,
we see that the maximal time scale
$t_m(r)$ in scheme $2$ is identical to that of scheme $1$,  and is given by
Eq.~(\ref{eq:tm}), since the same $\lambda_1$ again controls the long time decay of
$F_r(t)$ in scheme $2$ [see the second line of Eq.~(\ref{Fr2_asymp})]. Consequently, 
both schemes share the same $r_m^*$ that minimizes the time scale $t_m(r)$.

\section{Generalizations to non-Poissonian measurement protocols with a renewal structure}
\label{sec:general}

So far in this paper we have focused on Poissonian measurement protocol where in a small time $dt$, a 
measurement occurs with probability $r\, dt$, while with the complementary probability $(1-r\, dt)$ no 
measurement occurs. As mentioned before, this measurement process is Markovian since the probability of a 
measurement taking place at time $t$ does not depend on the previous history of the measurement. In other 
words, the measurement process is memoryless. In this case, the intervals between measurements are 
independent and identically distributed (IID) variables each drawn from $p(\tau)=r\, e^{-r\, \tau}$. The 
framework developed in this paper for this Poissonian measurement protocol can be easily generalised to 
the case where the measurement events are still uncorrelated as in the Poissonian case, but the waiting 
time between two successive measurements is distributed via an arbitrary distribution $p(\tau)$, not 
necessarily exponential. In this case the intervals between measurements are still IID variables, but 
with a general distriution $p(\tau)$. This is similar to the framework of the continuous-time random walk 
model. Such a renewal process with arbitrary $p(\tau)$ can not be generated by a memoryless measurement 
process as in the Poissonian case and in this sense it is sometimes referred to as `non-Markovian', 
even though the process still 
has a renewal structure. Only when $p(\tau)=r\, e^{-r\, \tau}$ is a pure exponential, the associated 
renewal process can be realized by a memoryless Markov measurement process. In this section, we
show how to compute the PDF $F_r(t)$ of the first detection time for such 
a non-Poissonian (and non-Markovian) measurement process characterized by IID intervals each drawn 
from a normalized PDF $p(\tau)$. This generalisation to non-Poissonian waiting time 
between measurements appeared in Ref.~\cite{KEZ21} and several results are known, in particular
for the return problem, i.e., scheme 2 in our notation.
We will see below that the survival probability formalism developed in this paper 
also generalizes quite easily and naturally to 
non-Poissonian case for both schemes $1$ and $2$.

For a general $p(\tau)$, the probability that there is no measurement in a time interval of duration 
$\tau$ is simply given by
\begin{equation}
q(\tau)= \int_{\tau}^{\infty} p(\tau)\, d\tau\, .
\label{qdef.gen1}
\end{equation}
This follows from the fact that for an interval after a measurement event to remain measurement free
up to time $\tau$, the next measurement must occur after time $\tau$, leading to Eq. (\ref{qdef.gen1}).
For the standard Poissonian case, using $p(\tau)= r\, e^{-r\, \tau}$, one gets from Eq. (\ref{qdef.gen1})
$q(\tau)= e^{-r \tau}$, as discussed in earlier sections. 

Our general formula for the survival probability $S_r(t)$ in Eq.~(\ref{surv.1}) still holds
for a general $p(\tau)$ not neceassarily exponential, except that the joint distribution
$P(n, \{\tau_i\}|t)$ now has a different expression. Using the IID property of the intervals,
we can express it as
\begin{equation}
P(n, \{\tau_i\}|t) =  \left[\prod_{k=1}^n p(\tau_k)\right]\,
q\left(t- \sum_{k=1}^n \tau_k\right)\, ,
\label{jpdf_product_gen}
\end{equation}
which replaces Eq.~(\ref{jpdf_product}) valid for 
$p(\tau) = r\, e^{-r\, \tau}$. Using again the delta function 
representation, Eq.~(\ref{jpdf_product_gen}) can be written in a more convenient form
\begin{equation}
P(n, \{\tau_i\}|t) =  \int_0^{\infty} d\tau_{\rm last}
\left[\prod_{k=1}^n p(\tau_k)\right]\, q(\tau_{\rm last})\, 
\delta\left(\tau_1+\tau_2+\ldots +\tau_n+ \tau_{\rm last}-t\right)\, .
\label{jpdf.gen1}
\end{equation}
This is the replacement of Eq.~(\ref{jpdf.1}). Taking Laplace transform with respect to $t$ and 
integrating over $\tau_{\rm last}$ replaces
Eq.~(\ref{jpdf_lap.1}) by
\begin{equation}
\int_0^{\infty} P(n, \{\tau_i\}|t)\, e^{-s\, t}\, dt =
\tilde{q}(s)\, \left[\prod_{k=1}^n p(\tau_k)\, e^{-s\, \tau_k}\right]\, ,
\label{jpdf_lap.gen1}
\end{equation}
where we have defined
\begin{equation}
\tilde{q}(s)= \int_0^{\infty} q(\tau)\, e^{-s\, \tau}\, d\tau\, .
\label{qlap_gen1}
\end{equation}
Note that from Eq.~(\ref{qdef.gen1}), it follows that $dq(\tau)/d\tau=- p(\tau)$. Taking Laplace
transform and using $q(0)=1$, one gets
\begin{equation}
\tilde{q}(s)= \frac{1-\tilde{p}(s)}{s}\, ,
\label{qlap_gen2}
\end{equation}
where 
\begin{equation}
\tilde{p}(s)= \int_0^{\infty} p(\tau)\, e^{-s\, \tau}\, d\tau\, .
\label{plap_gen1}
\end{equation}
One can also check the normalization of the joint PDF $P(n, \{\tau_i\}|t)$. By integrating over $\tau_i$'s
in Eq.~(\ref{jpdf_lap.gen1}), one gets
\begin{equation}
\int_0^{\infty} P(n|t)\, e^{-s\, t}\, dt= \tilde{q}(s)\, \left[\tilde{p}(s)\right]^n\, .
\label{pnt_lap_gen1}
\end{equation}
By summing over $n=0,1,2,\ldots$, and using the relation in Eq.~(\ref{qlap_gen2}), one easily checks
that 
\begin{equation}
\int_0^{\infty}\, \sum_{n=0}^{\infty} P(n|t)\, e^{-s\, t}\, dt= \frac{1}{s}\, ,
\label{norm_gen1}
\end{equation}
implying that $\sum_0^{\infty} P(n|t) = 1$ for all $t$.

Substituting Eq.~(\ref{jpdf.gen1}) into the general formula [Eq.~(\ref{surv.1})] and taking Laplace transform with respect 
to $t$, we arrive at the expression
\begin{equation}
\tilde{S}_r(s) =   \int_0^{\infty} dt\, e^{-s\, t}\, S_r(t) 
= \tilde{q}(s)+ \tilde{q}(s)\, \sum_{n=1}^{\infty}
\left[ \prod_{k=1}^n \int_0^{\infty} d\tau_k\,  p(\tau_k)\, e^{-s\, \tau_k}\,
S_r(n|\{\tau_i\})\right]\, .  
\label{surv_lap.gen1}
\end{equation}
This replaces the result in Eq.~(\ref{surv_lap.1}) valid for $p(\tau)=r\, e^{-r \tau}$.
Note that, as in Eq.~(\ref{surv_lap.1}), we have separated out the $n=0$ term from the sum.

The result in Eq.~(\ref{surv_lap.gen1}) is valid for general Hamiltonian and arbitrary $p(\tau)$.
To make further progress, we focus on a generic two level system, as in Sec.~\ref{sec:GE}.
Since $S_r(n|\{\tau_i\})$ does not depend on $p(\tau)$, it is given by the same expression 
as in Eq.~(\ref{eq:sn}), namely
\begin{equation}
S_r(n|\{\tau_i\})= f(\tau_1)\, \prod_{k=2}^n g(\tau_k) \, .
\label{eq:sn_gen}
\end{equation}
where the pair of functions $f(\tau)$ and $g(\tau)$ are still given by Eqs.~(\ref{eq:fdef})
and (\ref{eq:gdef}) respectively. Substituting Eq.~(\ref{eq:sn_gen}) in the general result
in Eq.~(\ref{surv_lap.gen1}) and performing the geometric sum over $n=1,2,\ldots$, 
we get the exact Laplace transform of the survival probability
\begin{equation}
\tilde{S}_r(s)= \tilde{q}(s)\, \left[ 1+ \frac{\tilde{U}(s)}{1-\tilde{V}(s)}\right]\, ,
\label{surv_lap_gen.2}
\end{equation}
where we have defined
\begin{eqnarray}
\tilde{U}(s) & =& \int_0^{\infty} d\tau\, p(\tau)\, f(\tau)\, e^{-s\, \tau} \label{U_def} \\
\tilde{V}(s) &= & \int_0^{\infty} d\tau\, p(\tau)\, g(\tau)\, e^{-s\, \tau}\, . \label{V_def}
\end{eqnarray}
Note that for the exponential case $p(\tau)= r\, e^{-r \tau}$, we have $\tilde{U}(s)= \tilde{f}(r+s)$
and $\tilde{V}(s)= \tilde{g}(r+s)$ and hence Eq.~(\ref{surv_lap_gen.2}) reduces to
Eq.~(\ref{eq:s_tilde_fg_laplace}). The Laplace transform of the first detection
probability is then given by
\begin{equation}
\tilde{F}_r(s)= 1- s\, \tilde{S}_r(s) 
= 1- s\, \tilde{q}(s)\, \left[1+ \frac{\tilde{U}(s)}{1-\tilde{V}(s)}\right]= 
1- \left[1-\tilde{p}(s)\right]\,\left[1+ \frac{\tilde{U}(s)}{1-\tilde{V}(s)}\right]\, .
\label{Frs_gen1}
\end{equation}
where we replaced $\tilde{q}(s)$ by its expression in Eq.~(\ref{qlap_gen2}).
For the Poissonian case $p(\tau)= r\, e^{-r\, \tau}$, Eq.~(\ref{Frs_gen1}) reduces to
Eq.~(\ref{eq:tildeF}). 

From this central expression in Eq.~(\ref{Frs_gen1}) valid for a generic two level system
and arbitrary $p(\tau)$, one can carry out exactly similar analysis
as in the previous sections to derive the general properties of the first detection probability and
its moments. We will skip some of the details here for the sake of brevity, but point out some
salient universal features. We consider the two schemes $1$ and $2$ separately below.

\subsection{First scheme}

We recall that in this scheme, the state of interest $|\psi_{\rm int}\rangle=|\psi_{-}\rangle$
differs from the initial state $|\psi_+\rangle$. Hence, in this case $f(\tau)=g(\tau)$ and
they are given in Eq.~(\ref{fg.S1}). Consequently, from Eqs. (\ref{U_def}) and (\ref{V_def}),
we get $\tilde{U}(s)=\tilde{V}(s)$. Plugging this in Eq.~(\ref{Frs_gen1}) gives
\begin{equation}
\tilde{F}_r(s)= \frac{\tilde{p}(s)- \tilde{V}(s)}{1- \tilde{V}(s)}\, ,
\label{Frs_S1}
\end{equation}
where $\tilde{V}(s)$ is given in Eq.~(\ref{V_def}) in terms of $g(\tau)$.
It is hard to invert this Laplace transform explicitly for general $p(\tau)$ and $g(\tau)$.
However, one can extract the asymptotic behavior of $F_r(t)$ for small and large $t$ as follows.

\vspace{0.4cm}

\noindent {\em Small $t$ behavior of $F_r(t)$.} To extract the small $t$ behavior of $F_r(t)$,
we consider the large $s$ behavior of its Laplace transform. Consider first
the Laplace transform $\tilde{p}(s)$ 
\begin{equation}
\tilde{p}(s) =  \int_0^{\infty} p(\tau)\, e^{-s\, \tau}\, d\tau 
= \frac{1}{s}\, \int_0^{\infty} d\tilde{\tau}\, e^{-\tilde{\tau}}\, p\left(\frac{\tilde{\tau}}{s}\right)\, ,
\label{lap_ps.S1}
\end{equation} 
where we rescaled $\tau=\tilde{\tau}/s$. Now, for large $s$, we can expand 
$p\left(\frac{\tilde{\tau}}{s}\right)$
in a Taylor series, assuming $p(\tau)$ is analytic near $\tau=0$ and then
perform the integral in Eq.~(\ref{lap_ps.S1}) term by term. This yields 
\begin{equation}
\tilde{p}(s)= \frac{p(0)}{s}+ \frac{p'(0)}{s^2}+ \frac{p''(0)}{s^3} + O\left(s^{-4}\right)\, .
\label{ps_exp.S1}
\end{equation}
Similarly, $\tilde{V}(s)$ in Eq.~(\ref{V_def}) can be expanded in a power series in $1/s$. Using $g(\tau)= 1-\sigma^2\, \tau^2$
as $\tau\to 0$ [see Eq.~(\ref{g_smallt})] where $\sigma^2$ is given in Eq.~(\ref{sigma2.1}), we get large $s$ 
\begin{equation}
\tilde{V}(s)= \frac{p(0)}{s}+ \frac{p'(0)}{s^2}+ \frac{\left(p''(0)- 2\sigma^2\right)}{s^3}+ 
O\left(s^{-4}\right)\, .
\label{Vs.S1}
\end{equation}
Substituting Eqs.~(\ref{ps_exp.S1}) and (\ref{Vs.S1}) in 
Eq.~(\ref{Frs_S1}), we get to leading order for large $s$
\begin{equation}
\tilde{F}_r(s) \approx \frac{2\, p(0)\, \sigma^2}{s^3}\, .
\label{Frs_bigs.S1}
\end{equation}
This implies that for small $t$
\begin{equation}
F_r(t) \approx p(0)\,\sigma^2\, t^2\, , \quad {\rm as}\quad t\to 0\, .
\label{Frt_small_S1}
\end{equation}
For the Poissonian case, using $p(0)=r$ one recovers the result in Eq.~(\ref{Frt_small.1}).
Thus, as long as $p(0)\neq 0$, one finds that the PDF of the first detection
probability exhibits a universal $t^2$ law for small $t$, independent of the details of $p(\tau)$
and also of the Hamiltonian of the two level system. The dependence on $p(\tau)$ and on $H$
appears only through the amplitude of this $t^2$ law, but the behavior $\sim t^2$ is universal
as long as $p(0)>0$. This universal $t^2$ law emerges from purely quantum dynamics that
dominates at early times.

\vspace{0.4cm}

\noindent {\em Large $t$ behavior of $F_r(t)$.} Extraction of the large $t$ behavior
of $F_r(t)$ from its Laplace transform in Eq.~(\ref{lap_ps.S1}) is more tricky.
If $p(\tau)$ decays exponentially or faster, one needs to investigate the poles
of $\tilde{F}_r(s)$ in the complex plane and in that case $F_r(t)$, for large $t$,
will decay exponentially, as was shown in the previous section for the Poissonian protocol.
However, if $p(\tau)$ has a power law tail for large $\tau$, e.g.,
\begin{equation}
p(\tau) \approx \frac{A}{\tau^{\mu+1}}\, , \quad {\rm with}\quad \mu>0\, ,
\label{pt_power_S1}
\end{equation}
its effect shows up as a singular term in the small $s$ expansion of the Laplace transform $\tilde{F}_r(s)$.
Consequently, one can extract the large $t$ behavior of $F_r(t)$ from the singular small $s$ behavior
of its Laplace transform. In fact, if $p(\tau)$ has a power law tail as in Eq.~(\ref{pt_power_S1})
for large $\tau$, its Laplace transform behave for small $s$ as
\begin{equation}
\tilde{p}(s)= 1 + \tilde{p}_{\rm reg}(s) + \tilde{p}_{\rm sing}(s)
\label{ps_smalls_S1}
\end{equation}
where $\tilde{p}_{\rm reg}(s)$ represents the regular part of the small $s$ expansion
and $\tilde{p}_{\rm sing}(s)$ is the leading singular term given by
\begin{equation}
\tilde{p}_{\rm sing}(s)\approx A_1\, s^\mu\, \quad {\rm where}\quad A_1= \Gamma(-\mu)\, A\, ,
\label{psing_S1}
\end{equation}
valid for all non-integer $\mu>0$ (for a simple proof see the appendix B of Ref.~\cite{EMZ06}). In Eq.~\eqref{psing_S1}, $\Gamma(x)$ denotes the Gamma function. 
For integer $\mu$, there will be additional multiplicative logarithm in the singular term~\cite{EMZ06},
but we omit the details here. 
Note that we have used $\tilde{p}(0)=1$ since $p(\tau)$ is normalized to unity.
Similarly, we can expand $\tilde{V}(s)$ into its regular and leading singular parts
for small $s$
\begin{equation}
\tilde{V}(s)= \tilde{V}(0) + \tilde{V}_{\rm reg}(s) + \tilde{V}_{\rm sing}(s)\, ,
\label{Vs_smalls_S1}
\end{equation}
where we will not need the detailed form of $\tilde{V}_{\rm sing}(s)$ as we will see shortly.
Substituting Eqs.~(\ref{ps_smalls_S1}) and (\ref{Vs_smalls_S1}) in Eq.~(\ref{Frs_S1}) and keeping only leading
order term for small $s$, we find that $\tilde{V}_{\rm sing}(s)$ drops out leaving 
\begin{equation}
\tilde{F}_r(s)= 1 + \tilde{F}_{\rm reg}(s) + \frac{A_1}{1-\tilde{V}(0)}\, s^\mu\, ,
\label{Frs_smalls_S1}
\end{equation}
where $\tilde{F}_{\rm reg}(s)$ represents the regular part of the small $s$ expansion of
$\tilde{F}_r(s)$.
From this singular small $s$ behavior in Eq.~(\ref{Frs_smalls_S1}), one can then read off the large $t$ behavior of $F_r(t)$ 
\begin{equation}
F_r(t) \approx \frac{A}{\left(1-\tilde{V}(0)\right)}\, \frac{1}{t^{\mu+1} }\,,
\label{Frt_larget_S1}
\end{equation}
where $A$ is the amplitude of the power law decay of $p(\tau)$ in Eq.~(\ref{pt_power_S1}). Thus,
the large $t$ behavior of $F_r(t)$ is completely governed by the large $\tau$ behavior
of $p(\tau)$, with the amplitude $A$ renormalized by the factor $1/(1- \tilde{V}(0))$.
The information about the Hamiltonian $H$ is contained only in 
$\tilde{V}(0)= \int_0^{\infty} d\tau\, p(\tau)\, g(\tau)$ through the function $g(\tau)$.
Thus the quantum Hamiltonian in the two level system only affects the amplitude of
the late time algebraic decay of $F_r(t)$ in Eq.~(\ref{Frt_larget_S1}). 

The mean detection time and the higher moments can also be computed easily from the Laplace transform $\tilde{F}_r(s)$ in Eq.~\eqref{Frs_S1}. For instance, if the first moment
$\langle \tau \rangle= \int_0^{\infty} \tau\, p(\tau)\, d\tau$ exists, then the mean detection time is given by the exact expression
\begin{equation}
\label{eq:tr_S1}
\bar{t}_r = \frac{\langle \tau \rangle}{1-\tilde{V}(0)}\, \quad {\rm where}\quad \tilde{V}(0)= \int_0^{\infty} p(\tau)\, g(\tau)\, d\tau\, .
\end{equation}

\subsection{Second scheme}

In this scheme, the state of interest $|\psi_{\rm int}\rangle=|\psi_{+}\rangle$
is the same as the initial state $|\psi_+\rangle$. In this case, we have
$f(\tau)+g(\tau)=1$ from Eq.~(\ref{eq:iden.1}) and hence $\tilde{f}(s)+\tilde{g}(s)=1/s$.
Adding Eqs.~(\ref{U_def}) and (\ref{V_def}), we then get 
\begin{equation}
\tilde{U}(s)+ \tilde{V}(s)= \tilde{p}(s)\, .
\label{UsVs_S2}
\end{equation}
Eliminating $\tilde{U}(s)$ in Eq.~(\ref{Frs_gen1}) using Eq.~(\ref{UsVs_S2}) gives
\begin{equation}
\tilde{F}_r(s)= 1- \frac{\left[1-\tilde{p}(s)\right]\, \left[ 1+\tilde{p}(s)-2\, 
\tilde{V}(s)\right]}{1-\tilde{V}(s)}\, . 
\label{Frs_S2}
\end{equation}

To extract the small $t$ behavior of $F_r(t)$, we expand $\tilde{F}_r(s)$ for large $s$ as in
scheme $1$ in the previous subsection. Omitting details, we get
\begin{equation}
\tilde{F}_r(s) \approx \tilde{V}(s)\approx \frac{p(0)}{s}\, ,
\label{Frs_larges_S2}
\end{equation}
where we used Eq.~(\ref{Vs.S1}) for the leading large $s$ expansion of $\tilde{V}(s)$.
The result in Eq.~(\ref{Frs_larges_S2}) implies 
\begin{equation}
F_r(t) \approx p(0)\, \quad {\rm as}\quad t\to 0 \, .
\label{Frt_smallt_S2}
\end{equation}
Thus this limiting small $t$ behavior is completely independent of the Hamiltonian $H$
and it depends only on $p(0)$, as long as $p(0)>0$. Note that for the Poissonian case
$p(\tau)=r\, e^{-r\, \tau}$ we have $p(0)=r$, thus recovering the result in Eq.~(\ref{smallr2.1}).

The large $t$ behavior of $F_r(t)$ can similarly be inferred from its Laplace transform
in Eq.~(\ref{Frs_S2}) following the same procedure as in scheme $1$ discussed in the
previous subsection. For the case when $p(\tau)= A/\tau^{\mu+1}$ for large $\tau$,  
we can again use the small $s$ expansion of $\tilde{p}(s)$ and $\tilde{V}(s)$
given respectively in Eqs.~(\ref{psing_S1}) and (\ref{Vs_smalls_S1}).
Substituting them in Eq.~(\ref{Frs_S2}) gives, for small $s$,
\begin{equation}
\tilde{F}_r(s) \approx 1 + \tilde{F}_{\rm reg}(s) +2\, p_{\rm sing}(s)
\approx 1 + \tilde{F}_{\rm reg}(s) +2\, A_1\, s^\mu\, ,
\label{Frs_smalls_S2}
\end{equation}
where $\tilde{F}_{\rm reg}(s)$ is the regular part of the small $s$ behavior of $\tilde{F}_r(s)$
and we have used the result $\tilde{p}_{\rm sing}(s)\approx A_1\, s^\mu$ 
from Eq.~(\ref{psing_S1}). From the singular small $s$ behavior in Eq.~(\ref{Frs_smalls_S2}),
one can then read off the large $t$ behavior of $F_r(t)$
\begin{equation}
F_r(t)\approx \frac{ 2\, A}{t^{\mu+1}}\, \quad {\rm as}\quad t\to \infty\, .
\label{Frt_larget_S2}
\end{equation}
Thus in both schemes, the leading large $t$ behavior of $F_r(t)$ has a universal power law tail
with exponent $\mu+1$ that does not depend on the Hamiltonian $H$. In scheme $1$, the
dependence on the Hamiltonian shows up in the amplitude through $\tilde{V}(0)$ in
Eq.~(\ref{Frt_larget_S1}). In contrast, in scheme 2, even the amplitude
is independent of $H$, making the large $t$ tail of $F_r(t)$ completely universal.

The mean detection time in scheme $2$ can also be computed explicitly from the Laplace transform $\tilde{F}_r(s)$ in Eq.~\eqref{Frs_S2}. As long as $\langle \tau \rangle= \int_0^{\infty} \tau\, p(\tau)\, d\tau$ exists, the mean first detection time is given by 
\begin{equation}
\label{eq:tr_s2}
\bar{t}_r = 2\, \langle \tau \rangle\, .
\end{equation}
As mentioned earlier, this result 
in Eq.~\eqref{eq:tr_s2} was derived in Refs.~\cite{KEZ21, ZBK21} 
using a different method
and the factor $2$ was shown to be the dimension $2$ of the Hilbert space for
non-degenerate two level systems. A more general result for $N$ level systems states
$\bar{t}_r= N\, \langle \tau \rangle$ where $N$ is the dimension of
the Hilbert space in non-degenerate systems. This general result was first
discovered for stroboscopic protocol in Ref.~\cite{GVWW13}, and was subsequently generalised
to other random protocols in Ref.~\cite{KEZ21}.
For the stroboscopic protocol, the relation in Eq. (\ref{eq:tr_s2}) has 
been recently measured experimentally using an IBM quantum computer~\cite{TZ22}.

\section{Conclusions and Outlook}
\label{sec:conc}

To summarize, in this paper we have studied `quantum resetting' purely via random projective 
measurements. In our system, `restart' is not implemented by classical resetting of a 
quantum state, but rather the projective measurements at random times perform the resetting.  
Our main object of interest is the probability distribution $F_r(t)$ of the first detection 
time of a specified quantum state of interest. We have presented a general framework to 
compute $F_r(t)$ for arbitrary measurement protocols. We then focused on the protocol where 
the measurement events are independent, with the waiting time between successive 
measurements drawn independently from a general PDF $p(\tau)$. The case $p(\tau)= r\, e^{-r 
\tau}$ corresponds to the Poissonian protocol where in a small time $dt$, a measurement 
occurs with probability $r\, dt$ and with the complementary probability $(1-r\, dt)$ no 
measurement occurs. We have shown how our general framework yields exact results for 
$F_r(t)$ in an arbitrary two level system. It turns out that the result depends crucially on 
the detection schemes involved.  We considered two complementary schemes: (1) where the 
state of interest coincides with and (2) differs from the initial state. Interestingly, the 
mean first detection time, as a function of the measurement rate $r$, exhibits markedly 
different behaviors in the two schemes. In scheme $1$, the PDF $F_r(t)$ vanishes as $t\to 0$ 
universally as $F_r(t)\sim t^2$ (independent of $p(\tau)$ as long as $p(0)\neq 0$), while in 
scheme $2$, $F_r(t)$ approaches a constant $p(0)$ as $t\to 0$. Also, the mean detection 
time in scheme $1$ is a non-monotonic function of $r$ with a single minimum at an optimal 
value $r^*$, while in scheme $2$ it is a monotonically decreasing function of $r$, 
signalling the absence of a finite optimal value. These general predictions are verified via 
explicit computation in the Jaynes-Cummings model of light-matter interaction. We also 
generalised our results to non-Poissonian measurement protocols with a renewal structure 
where the intervals between successive independent measurements are distributed via a 
general distribution $p(\tau)$. We showed that the short time behavior of $F_r(t)\sim p(0)\, 
t^2$ in scheme $1$ is universal and this universalilty is rooted in purely 
quantum dynamics that dominates at early times.

In this paper, we derived explicit results for the statistics of the first detection
time in two complementary detection schemes for
a generic two level system subjected to 
Poissonian (and non-Poissonian) random measurement protocol. It would be interesting to see 
if our results, especially for scheme $1$, can be extended to systems with more than $2$ states that are subjected to 
these random measurement protocols. 

The general framework presented here, for Poissonian 
and non-Poissonian measurement protocols, may be useful in other quantum measurement 
problems of current 
interest. For example, a very interesting area of current research concerns measurement 
induced phase transitions in quantum systems~\cite{LCF19,SRN19,GH20,ZGWGHP20, RCGG20, 
SSLT22}. It would be interesting to explore if the formalism developed in this paper can be 
extended to study the dynamics of entanglement entropy under repeated measurements using  
Poissonian/non-Poissonian random protocols.

\section{Acknowledgements}

We thank Eli Barkai for useful discussions. M.~K. would like to acknowledge support from the Project 6004-1 of the Indo-French Centre for the Promotion of 
Advanced Research (IFCPAR), Ramanujan Fellowship (SB/S2/RJN114/2016), SERB Early Career Research Award 
(ECR/2018/002085) and SERB Matrics Grant (MTR/2019/001101) from the Science and Engineering Research Board (SERB), 
Department of Science and Technology, Government of India. M. K. acknowledges support from the Department of 
Atomic Energy, Government of India, under Project No. RTI4001. 
M.~K. thanks the hospitality of LPENS (Paris), LPTHE (Paris) and LPTMS (Paris-Saclay) during several visits.  
M.~K. acknowledges the support from the Science and
Engineering Research Board (SERB, government of India),
under the VAJRA faculty scheme (No. VJR/2019/000079).
S.~N.~M. acknowledges the support from the Science and
Engineering Research Board (SERB, government of India),
under the VAJRA faculty scheme (No. VJR/2017/000110).

\appendix

\section{Exact solvability and quantum time dynamics of the Jaynes-Cummings Hamiltonian}
\label{app_JC}

\subsection{Integrability}
\label{Int_JC}

In this appendix, for the sake of completeness we briefly review the well known 
Jaynes-Cummings (JC) Hamiltonian~\cite{GSA13,SZ97,C99,HR06} and discuss its exact solvability. The  Jaynes-Cummings Hamiltonian is given by 
\begin{equation}
H = \omega_q S^z + \omega_c a^{\dagger} a + g [S^{+} a + a^{\dagger} S^-],
\label{eq:app_JCHam}
\end{equation}
where $\omega_q$ is the qubit frequency, $\omega_c$ is the cavity frequency and $g$ is the 
light-matter coupling. Here $a / a^{\dagger}$ are annihilation/creation operators of bosons and 
$S^{\pm}, S^z$ are the Pauli spin operators. The model [Eq.~\eqref{eq:app_JCHam}] is defined on a direct product of two spaces, i.e., a two level Hilbert space (spanned by $\mid \uparrow \rangle, \mid\downarrow \rangle$) and a single mode cavity Hilbert space (spanned by $\big| n \rangle$ where $n =  0,1,2,3 \dots$. The algebra for operators is given by 
\begin{equation}
[a,a^{\dagger}] = 1,\,  [S^z, S^{\pm}] =  \pm S^{\pm}, \, [S^+, S^-] = 2S^z. 
\label{eq_app_com}
\end{equation}
Let us define an excitation operator $\hat{N}$ as 
\begin{equation}
\label{eq:Nex}
\hat{N}  = a^{\dagger} a + S^+ S^-. 
\end{equation}
It is easy to check that [using Eq.~\eqref{eq_app_com}] the operator $\hat{N}$ and $H$ commute with each other, i.e., 
\begin{equation}
[H,\hat N] = 0.
\end{equation}
Also, from Eq.~\eqref{eq:Nex}, one can show
\begin{equation}
\label{eq:N_sec}
\hat{N}  \mid \downarrow, n \rangle  = n  \mid \downarrow, n \rangle, \, \quad 
\hat{N}  \mid \uparrow, n-1 \rangle  = n  \mid \uparrow, n-1 \rangle .
\end{equation} 
From Eq.~\eqref{eq:N_sec}, one concludes that for a given excitation sector labelled by $n$, the Hilbert space is 
spanned by only two states $\mid \downarrow, n \rangle$ and $\mid \uparrow, n-1 \rangle$. The Hamiltonian does not 
couple spaces spanned by different sectors. Therefore, in a 
given excitation sector labelled by $n$, the JC Hamiltonian can be represented by a $2\times 2$ matrix
\begin{eqnarray}
H_n &=& \begin{bmatrix}
\label{eq:hn}
 \langle \downarrow, n | H | \downarrow, n \rangle  & \langle \downarrow, n | H | \uparrow, n-1 \rangle \\
 \langle \uparrow, n-1 | H | \downarrow, n \rangle & \langle \uparrow, n-1 | H | \uparrow, n-1 \rangle \end{bmatrix}\nonumber \\
 &=& 
 \begin{bmatrix}
 -\frac{\omega_q}{2} + \omega_c n & g \sqrt{n} \\
  g \sqrt{n} &  \frac{\omega_q}{2} + \omega_c (n-1) \end{bmatrix}.
\end{eqnarray} 
The eigenvectors and eigenvalues of $H_n$ [Eq.~\eqref{eq:hn}] are given by 
\begin{eqnarray}
\label{eq:vec}
| n, + \rangle &=& \cos(\theta_n) \mid \downarrow, n \rangle + \sin(\theta_n)  \mid \uparrow, n-1 \rangle \, ,
\nonumber \\
| n, - \rangle &=&  - \sin(\theta_n) \mid \downarrow, n \rangle + \cos(\theta_n) \mid \uparrow, n-1 \rangle  \, ,
\end{eqnarray}
and 
\begin{equation}
\label{eq:val}
E_{n_{\pm}}  =  \omega_c \bigg(n - \frac{1}{2} \bigg) \pm  \frac{1}{2}\sqrt{(\omega_c - \omega_q)^2 + 4g^2 n}. 
\end{equation}
In Eq.~\eqref{eq:vec}, $\theta_n$ (mixing angle) is defined to be
\begin{equation}
\label{eq_app_theta}
\theta_n  = \frac{1}{2}\tan^{-1} \bigg[\frac{2g \sqrt{n}}{\omega_c - \omega_q} \bigg].
\end{equation}
Therefore, in a given sector the JC Hamiltonian [Eq.~\eqref{eq:app_JCHam}] is exactly solvable (integrable). 

\subsection{Initial value problem}
\label{Initial_JC}

After establishing the integrability/exact solvability, we will now discuss the initial value problem, i.e.,
how a state evolves in time starting from an initial state.  Let us choose the following initial condition
\begin{equation}
\label{eq:JC_init_app}
 | \psi(0) \rangle = \mid \uparrow, n-1 \rangle .
\end{equation}
We will now find out how the initial condition in Eq.~\eqref{eq:JC_init_app} evolves when subjected to the JC Hamiltonian [Eq.~\eqref{eq:app_JCHam}]. In other words, we need to evaluate
\begin{equation}
\label{eq:se_app}
| \psi(t) \rangle = e^{-i H t} | \psi(0) \rangle
= e^{-i H t}  \mid \uparrow, n-1 \rangle .
\end{equation}
To simplify things further, we will stay in the resonant limit, $\omega_c = \omega_q$ which implies $\theta_n = \pi/4$ [Eq.~\eqref{eq_app_theta}].  In this resonant limit,  Eq.~\eqref{eq:vec} and Eq.~\eqref{eq:val} simplify to 
\begin{eqnarray}
\label{eq:vec_res}
| n, + \rangle &=& \frac{1}{\sqrt{2}} \Big[   \mid \uparrow, n-1 \rangle  + \mid \downarrow, n \rangle  \Big] \, , \nonumber \\
| n, - \rangle &=&   \frac{1}{\sqrt{2}} \Big[   \mid \uparrow, n-1 \rangle  - \mid \downarrow, n \rangle  \Big] \, ,
\end{eqnarray}
and 
\begin{equation}
\label{eq:val_res}
E_{n_{\pm}}  =  \omega_c \bigg(n - \frac{1}{2} \bigg) \pm  g\sqrt{n}. 
\end{equation}
Inverting Eq.~\eqref{eq:vec_res}, we get
\begin{eqnarray}
\label{eq:vec_res_inv}
\mid \uparrow, n-1 \rangle &=& \frac{1}{\sqrt{2}}  \Big[ | n, + \rangle +  | n, - \rangle  \Big] \, ,\nonumber \\
\mid \downarrow, n \rangle &=& \frac{1}{\sqrt{2}}  \Big[ | n, + \rangle -  | n, - \rangle  \Big] .
\end{eqnarray}
Using Eq.~\eqref{eq:vec_res_inv}, Eq.~\eqref{eq:se_app} then becomes
\begin{eqnarray}
\label{eq:se_app_simp}
| \psi(t) \rangle = e^{-i H t}  \mid \uparrow, n-1 \rangle  
&=&  \frac{1}{\sqrt{2}} \Big[ e^{-i E_{n_{+}} t} \mid n, + \rangle +   
e^{-i E_{n_{-}} t} \mid n, - \rangle  \Big] .
\end{eqnarray}
Substituting the the expressions for $E_{n_{\pm}}$ from Eq.~\eqref{eq:val_res} in Eq.~\eqref{eq:se_app_simp}, we get 
\begin{equation}
\label{eq:se_app_simp_final}
| \psi(t) \rangle = e^{-i\omega_c \big(n- \frac{1}{2} \big)t}  \big[ - i \sin(g\sqrt{n} t)  \mid \downarrow, n \rangle 
+ \cos(g\sqrt{n} t) \mid \uparrow, n-1 \rangle \big]\, . 
\end{equation}
From Eq.~\eqref{eq:se_app_simp_final}, we can see that the 
probability that the system stays in excited state is
\begin{equation}
\label{eq:pe}
P_e(t) = \langle  \psi(t) | \uparrow \rangle \langle \uparrow    | \psi(t) \rangle  =  \cos^2(g\sqrt{n} t).
\end{equation}
and similarly the probability that the system stays in ground state is given by 
\begin{equation}
P_g(t) = \langle  \psi(t) |\downarrow \rangle \langle \downarrow    | \psi(t) \rangle  =  \sin^2(g\sqrt{n} t).
\end{equation}
Similar to Eq.~\eqref{eq:se_app_simp}, if we had started with an initial state $ | \psi(0) \rangle = \mid \downarrow, n \rangle $ we get
\begin{eqnarray}
\label{eq:se_app_simp_down}
| \psi(t) \rangle 
&=& \frac{1}{\sqrt{2}} \big[ e^{-i E_{n_{+}} t} \mid n, + \rangle 
-  e^{-i E_{n_{-}} t} \mid n, - \rangle  \big] \nonumber \\ &=& 
e^{-i\omega_c \big(n- \frac{1}{2} \big)t}  \big[ - i \sin(g\sqrt{n} t)  
\mid \uparrow, n \rangle  + \cos(g\sqrt{n} t) \mid \downarrow, n-1 \rangle \big]\, . 
\end{eqnarray}\
Using similar analysis, one can solve the JC model for any arbitrary initial condition by exploiting the fact that the Hamiltonian takes a block diagonal form consisting of $2 \times 2$ sectors. The results presented in this appendix were used in Sec.~\ref{sec:JC_mfd} for computing the first detection probability for the JC system.

\end{document}